\documentstyle[preprint,tighten,floats,twoside,eqsecnum,aps,epsf]{revtex}

\title{
\hfill\hbox{\normalsize\rm HD-THEP-96-31}\\
\hfill\hbox{\normalsize\rm HD-TVP-96-8}\\ \vspace{3cm}
Vector meson leptoproduction\\ and nonperturbative gluon fluctuations in QCD}

\author{H.G. Dosch, T. Gousset\thanks{Supported by the Federal Ministry of Education, 
Science, Research and Technology (BMBF) under grant no. 06 HD 742}, 
G. Kulzinger\thanks{Supported by the Deutsche Forschungsgemeinschaft under grant
no. GRK 216/1-96}, H.J. Pirner\\\vspace{.5cm}
{\normalsize\it Institut f\"ur Theoretische Physik der Universit\"at Heidelberg,\\
Philosophenweg 16 \& 19, D69120 Heidelberg}\vspace{1.5cm}\\}

\def\slash{\llap{/}}
\newcommand{\fips}[1]{\epsffile{#1.ps}}

\begin{document}

\maketitle

\begin{abstract}
\noindent
We present a nonperturbative QCD calculation of diffractive vector meson 
production in virtual photon nucleon scattering at high energy. We use the 
nonperturbative model of the stochastic QCD vacuum which yields linear 
confinement and makes specific predictions for the dependence of high-energy 
scattering cross sections on the hadron size. Using light cone wave functions of 
the photon and vector mesons, we calculate electroproduction cross sections 
for $\rho$, $\omega$, $\phi$ and $J/\psi$. We emphasize the behavior of 
specific observables such as the ratio of longitudinal to transverse production 
cross section and the $t$-dependence of the differential cross section.  
\end{abstract}
\thispagestyle{empty}
\newpage

\section{Introduction}

Exclusive vector meson production by real and virtual photons is a good probe 
to investigate the physics of diffractive scattering. Whenever the coherence 
length of the photon is larger than the proton radius, it is preferable to study 
the process in the proton rest frame or in the center of mass frame where the 
virtual photon can be considered as a hadronic system composed of partons. 
In this case the photon-hadron interaction is then in many respects similar to 
hadron-hadron collision. In addition, it offers the possibility to vary the 
polarization and virtuality of the photon and thereby manipulate the light 
cone wave function of the incoming state. ``The experimentalist can make 
hadrons of arbitrary sizes.''

In our approach, we attack the problem as a genuine nonpertubative one. We 
use the model of the stochastic vacuum~\cite{dos87} which has been adapted 
to high-energy hadron-hadron scattering in Ref.~\cite{dos94}, applying the 
general scheme developped by Nachtmann~\cite{nac91} for the separation of 
the large energy scale from the small scale of momentum transfer. The model 
of the stochastic vacuum gives satisfactory results both in low and high energy 
physics. It yields a rather simple geometrical picture for a single gluonic flux 
tube~\cite{rue95}. The same mechanism of nonperturbative gluon fluctuations 
which leads to confinement also generates an interaction of the strings in 
colliding hadrons. In this picture hadron-hadron scattering cannot be 
constructed from quark-quark scattering since the string-string interaction 
plays an important role. It leads to cross sections which are determined by the 
transverse extensions of the interacting hadrons. In addition to the forward 
scattering amplitudes, the model provides the $t$-dependence of the cross 
section explaining the phenomenologically observed~\cite{pov87} correlation 
between elastic slopes and total cross sections.

Soft electroproduction on the nucleon can be calculated along the same 
lines as hadron-nucleon scattering using a model wave function for the photon. 
At small $Q^2$, the photon is of hadronic size and large-distance-physics as in 
hadron-hadron scattering should apply. In another paper~\cite{dos96}, we 
construct the wave function of the photon as a superposition of vector meson 
states and calculate the production cross section of $\rho$, $\rho'$ and $\rho''$. 
This approach is limited to virtualities $Q^2\lesssim2\,$GeV$^2$ and the 
unknown couplings of the electromagnetic current to the $\rho'$ and $\rho''$ 
introduce new parameters. In this paper, we explore the possibility to represent 
the incoming photon as a $q\bar{q}$ state. With increasing $Q^2$, the 
transverse extension of the $q\bar{q}$ dipole diminishes in a way that depends 
on the polarization of the virtual photon. We shall demonstrate how this 
mechanism shows up phenomenologically. This allows us to study the 
transition from large to short-distance-dominated-processes. The 
cross section for transversely polarized photons has a large non perturbative 
part, because endpoints at momentum fraction $z=0$ and $z=1$ in the photon 
wave function do not select a $q\bar{q}$ system of small transverse separation. 
Therefore our large distance mechanism is important here, too.

In our model the length of the string connecting the valence quarks in the hadron 
turns out to be very important. This length depends on the light cone wave function 
of the hadron. There is yet little knowledge about the physics determining the 
light cone Hamiltonian in nonperturbative QCD. So at the moment these wave 
functions and their integrated distribution amplitudes fulfill mainly a 
phenomenological task to parametrize the valence quark content of the hadron. 
Although, the exact value of cross section depends in our model on the detailed 
form of the wave function, the $Q^2$-behavior of specific observables such as 
ratios of longitudinal to transverse vector meson production or elastic slopes are 
likely to provide a good test for the string picture inherent in our model.

Electroproduction of vector mesons has also been discussed within a soft pomeron 
framework in Ref.~\cite{don87,don95}. In this model, transverse sizes of hadrons 
only play a marginal role because, on the one hand, hadron scattering can 
systematically be reduced to quark scattering through the property of quark 
additivity of the model and, on the other hand, the vector meson wave functions 
are assumed to be wider than the distance of the quarks in the virtual photon 
and are thus replaced by their value at the origin. These assumptions are 
phenomenologically tenable if one further assumes that the quark-pomeron 
coupling is flavor dependent. At intermediate $Q^2=1-10\,$GeV$^2$~\cite{don87}, 
a pomeron form factor is used for far-off-shell quark legs. At larger values of 
$Q^2$~\cite{don95}, nonperturbative two-gluon exchange is applied which leads 
to color singlet cancellation at small $q\bar{q}$-dipole size.

In a series of papers~\cite{kop93}, a perturbative two-gluon exchange model 
extended to include nonperturbative effects via the gluon distribution in the 
proton has been developped to evaluate vector meson production. This approach 
has been further refined to incorporate a BFKL-like evolution to accomodate both 
energy and $Q^2$-dependence. Dipole scattering is the basis of this framework. 
In the following paper we are treating the photon and vector meson in a similar 
way. The main difference between our approaches lies in the reaction mechanism 
for soft diffraction. 

The importance of the gluon distribution as a necessary part of hard diffraction
has been advocated in Ref.~\cite{rys93,bro94}. These authors have limited the range 
of applicability of their perturbative calculations to reactions where a large 
transverse momentum scale rules the exchanges, i.e. to heavy quark production 
such as $J/\psi$-production~\cite{rys93} or at large $Q^2$ 
($Q^2\ge 10\,$GeV$^2$~\cite{bro94}).
  
HERA has opened up new possibilities to enlarge the energy and $Q^2$ range. 
Recent ZEUS and H1 data indicate that at large $Q^2$ the cross section may rise 
more steeply than expected from soft pomeron exchange. A possible explanation 
in the language used above~\cite{mue94} is that the evolution of the wavefunction 
at higher energy and $Q^2$ gives rise to more and more dipoles inside the hadron. 
This evolution would then be a prerequisite to discuss the energy dependence of the 
virtual photon cross section with increasing resolution. This  phenomenon of ``hard'' 
pomeron exchange will not be addressed in the following study which deals with soft 
pomeron physics at a fixed energy.
\medskip

\begin{figure}[ht]
\vskip 2mm
$$
\fips{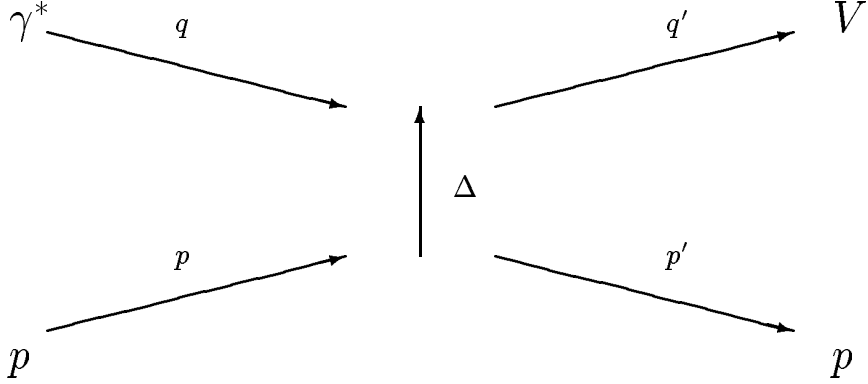}
$$
\caption{Kinematics of the reaction $\gamma^* +p\to V+p$.}
\label{kinematics}
\end{figure}

The kinematics are defined in Fig.~\ref{kinematics}. We denote the initial photon 
4-momentum with $q$, the initial nucleon momentum with $p$, and the equivalent 
final states with $q'$ and $p'$. $\Delta=q'-q$ is the momentum transfer and the 
independent Lorentz invariants are 
\begin{eqnarray*}
s&=&(p+q)^2,\\
t&=&\Delta^2=(q'-q)^2,\\
Q^2&=&-q^2.
\end{eqnarray*}
We are interested in soft reactions, i.e. $|t|<1\,$GeV$^2$, at high energy, 
$s\gg Q^2$ and $s\gg M_p^2$ (e.g. $s>100\,$GeV$^2$). In this domain, 
$x_{\rm B}=Q^2/2p.q$ is small, e.g. $x_{\rm B}<0.1$. 

To be specific, we use the center-of-mass frame where the photon momentum 
points along the $z$-axis. Then the absolute sizes of the 3-momenta are given as
\begin{eqnarray*}
|p|=|q|&=&{\sqrt{s}\over2}+{Q^2-M_p^2\over2\sqrt{s}}+O(s^{-3/2}),\\
|p'|=|q'|&=&{\sqrt{s}\over2}-{M_V^2+M_p^2\over2\sqrt{s}}+O(s^{-3/2}).
\end{eqnarray*}
In this frame the vector meson emerges with a small transverse momentum 
$|{\bf q}'_T|=\Delta_T\approx\theta\sqrt{s}/2<1\,$GeV. Unlike 
in the elastic scattering case, the center of mass momentum varies in this 
reaction 
$$
\delta=|p|-|p'|\approx {Q^2+M_V^2\over2\sqrt{s}}.
$$
This implies that the momentum transfer $\Delta$ has a time component
$\Delta^0\approx\delta$, besides the space components 
$\Delta^z\!\approx\!-\delta\!-\!\sqrt{s}\,\theta^2/4$ and 
$\Delta_T\!\approx\!\sqrt{s}\,\theta/2$. Let us notice for completness that the 
square of $\Delta$ is $t=\Delta^2\approx -\Delta_T^2+t_0$ with 
$t_0=-M_p^2(Q^2+M_V^2)^2/s^2$. In the high energy limit $s\gg Q^2+M_V^2$ 
the space components dominate. When we demand in addition a finite transfer, 
i.e. a small scattering angle $\theta/2=O(1/\sqrt{s})$, the transverse component 
is the leading component of $\Delta$. In the following we shall therefore neglect 
all component besides ${\bf\Delta}_T$.
\medskip

The outline of the paper is as follows. Section II gives a very short and 
non-technical description of the model of the stochastic vacuum and describes 
its application to high-energy scattering. Section III deals with the specific 
features of electroproduction, i.e. the photon and vector meson wave functions, 
which are used in the evaluation of the longitudinal and transverse cross 
sections. Sections IV contains the numerical results for $\rho$, $\phi$, $J/\psi$ 
integrated and differential cross sections as functions of $Q^2$. As far as 
possible these results are compared to experiment. Section V concludes with a 
discussion of the results.

\section{High-energy elastic scattering in the stochastic vacuum}\label{scattering}

\subsection{The model of the stochastic vacuum}

The model of the stochastic vacuum is based on the asumption that the 
contributions of the slowly varying gluon fields in an infrared regular QCD can 
be approximated by a simple stochastic process (for a review see 
Ref.~\cite{dos94b}). Already the assumption that this process has a converging 
cluster expansion leads to linear confinement in a non-Abelian gauge theory. 
As usual, approximations to a quantum field theory in the functional approach 
are more safely made in an Euclidean rather than Minkowskian field theory. It 
turns out however that in high-energy scattering there is no feasible way to 
continue Greens functions from the Euclidean to the Minkowskian world and we 
therefore have to formulate the model in the Minkowski continuum. This seems 
at first sight more dramatic than it turns out to be finally, since at the end we 
have to evaluate the relevant quantities only at spacelike Euclidean 
distances, i.e we can take these quantities from an Euclidean field theory.

In order to define gauge invariant correlators we introduce the modified 
gluon field strength $F_{\mu\nu}(x,\omega)$ which is obtained from the field 
strength at point $x$ by parallel transporting the colour content to the point 
$\omega$
$$
F_{\mu\nu}(x,\omega)=\Phi^{-1}(x,\omega)F_{\mu\nu}(x)\Phi(x,\omega),
$$
with $\Phi(x,\omega)=P\exp[-ig\int_\omega^x Adz]$.

Assuming that the main features of the correlator
$\langle F_{\mu\nu}(x,\omega)F_{\rho\sigma}(y,\omega)\rangle$ do not 
depend crucially on the choice of the reference point $\omega$ we obtain 
for the dependence on $z=x-y$ the most general form 
\begin{eqnarray}\label{correlator}
\langle g^2F^c_{\mu\nu}(x,\omega)F^d_{\rho\sigma}(y,\omega)\rangle_A
&=&{\delta^{cd}\over N_c^2-1}{1\over12}\langle g^2FF\rangle \Big\{\kappa
(\eta_{\mu\rho}\eta_{\nu\sigma}-\eta_{\mu\sigma}\eta_{\nu\rho})D(z^2/a^2)\\
&&+(1-\kappa){1\over2}\Big[\partial_{\mu}(z_{\rho}\eta_{\nu\sigma}
-z_{\sigma}\eta_{\nu\rho})+\partial_{\nu}(z_{\sigma}\eta_{\mu\rho}
-z_{\rho}\eta_{\mu\sigma})\Big]D_1(z^2/a^2)\Big\}.\nonumber
\end{eqnarray}
The correlator $D$ is typical for a non-Abelian gauge theory (or an Abelian theory 
with monopoles) since the homogeneous Maxwell equations 
$$
 \epsilon^{\mu\nu\rho\sigma}\partial_\nu F_{\rho\sigma}=0
$$
allow only the tensor structure proportinal to $D_1$, hence $\kappa=0$ in an Abelian 
theory without monopoles. 

In a Gaussian model, where all higher cumulants in the linked cluster 
expansion~\cite{kam76} are neglected, we obtain a relation between the slope of 
the static quark-antiquark potential and the typically non-Abelian correlator $D$
$$
\sigma=\kappa{\pi\over144}\langle g^2FF\rangle a^2\int_0^{+\infty}du\,D(-u).
$$

The choice of phenomenological parameters will be given in Sec.~\ref{hadron-hadron}.

\subsection{Scattering of two color singlet dipoles}\label{dipole-scattering}

The high-energy scattering of two color singlet dipoles  $q_1 \bar{q}_1$
and $q_2 \bar{q}_2$ can be treated analogously to the situation of 
heavy quarks encountered in the Wilson area law. The relativistic quarks and 
antiquarks move along two opposite straight line trajectories on the light cone. 
In order to apply the model of the stochastic vacuum to high-energy hadron hadron 
scattering we adopt the method of Ref.~\cite{nac91}. In this approach the problem 
is first considered as the scattering of quarks in an external colour field which is 
solved for fast moving quarks by the leading term of an eikonal expansion, i.e. the 
quark picks up the eikonal phase
$$
V=\exp[-ig\int_\Gamma A dz]
$$
where $\Gamma$ is the classical path of the quark. 

\begin{figure}
\vskip 2mm
$$
\fips{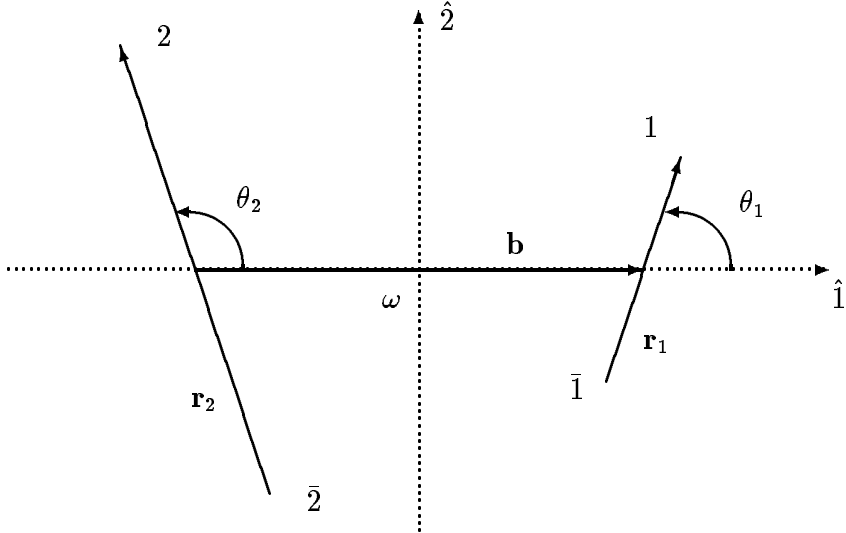}
$$
\caption{Configuration of the two interacting loops in the transverse plane. With our 
choice of frame, the loops 1 and 2 lie, in the $(x^0,x^3)$-plane, on the lines 
$x^0=x^3$ and $x^0=-x^3$ respectively.}
\label{loop-loop}
\end{figure}

This phase is manifestly gauge dependent, but if we consider a fast moving dipole, 
i.e. a quark and an antiquark moving on parallel lightlike trajectories connected 
by a Schwinger string, then we have to evaluate rather a Wilson loop 
$$
W=\exp[-ig\int_{\partial S} A dz],
$$
than the path integral above. The open ends of the $q$ and $\bar{q}$ trajectories 
in dipole 1 and dipole 2 are closed by small transverse lines yielding two loops, 
$\partial S_1$ and $\partial S_2$, which have transverse extensions according to 
the lengths of the dipoles $r_1$ and $r_2$. The dipoles are positioned relative to 
each other with a given impact parameter $b$. The loop-loop interaction amplitude 
for this system of dipoles is calculated in Ref.~\cite{dos94} 
$$
J({\bf x}_1,{\bf x}_{\bar 1},{\bf x}_2,{\bf x}_{\bar 2})=
\left\langle{1\over N_C}{\rm tr}\left(W_1({\bf x}_1,{\bf x}_{\bar 1})-1\right)
{1\over N_C}{\rm tr}\left(W_2({\bf x}_2,{\bf x}_{\bar 2})-1\right)\right\rangle_A,
$$
where the bold faced vectors ${\bf x}_i=(x_i^1,x_i^2)$ denote the two-dimensional
positions of quarks $i=1,2$ in the transverse plane ($\bar{\imath}$ refers to the 
corresponding antiquarks). The geometry of the loops is shown in Fig.~\ref{loop-loop}. 

\begin{figure}
\vskip 2mm
$$
\fips{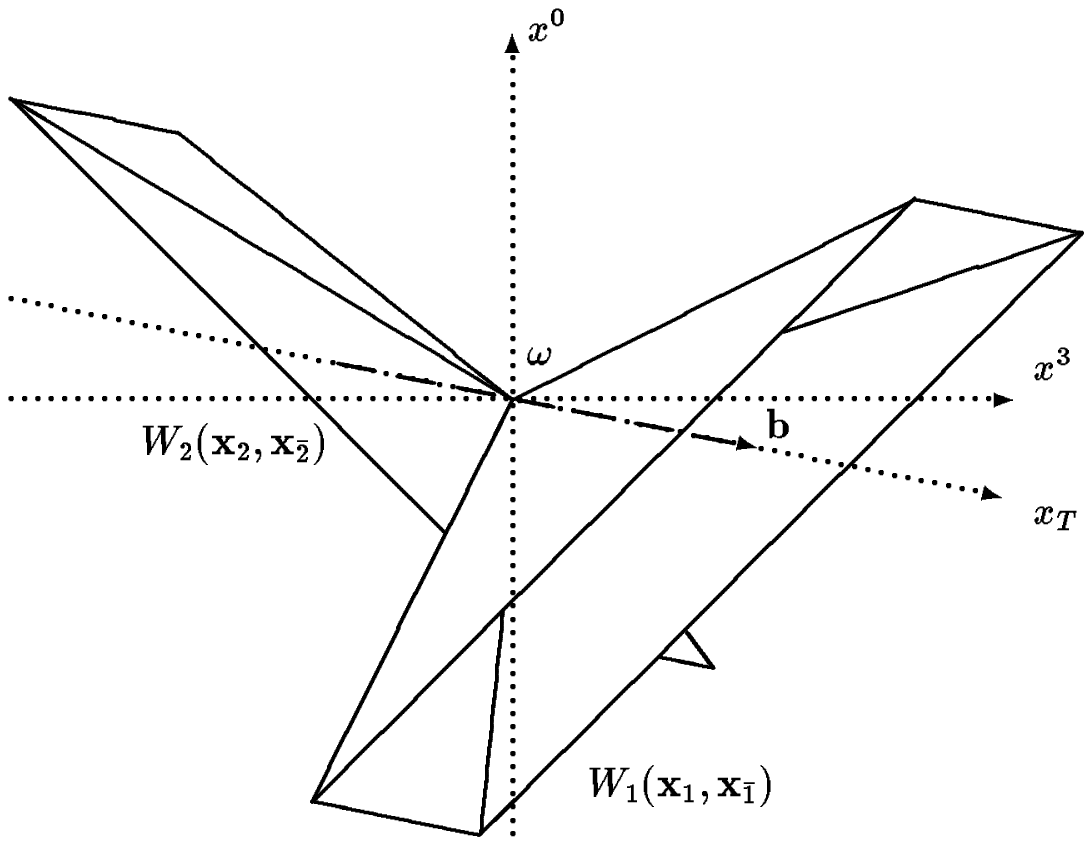}
$$
\vskip 3mm
$$
\fips{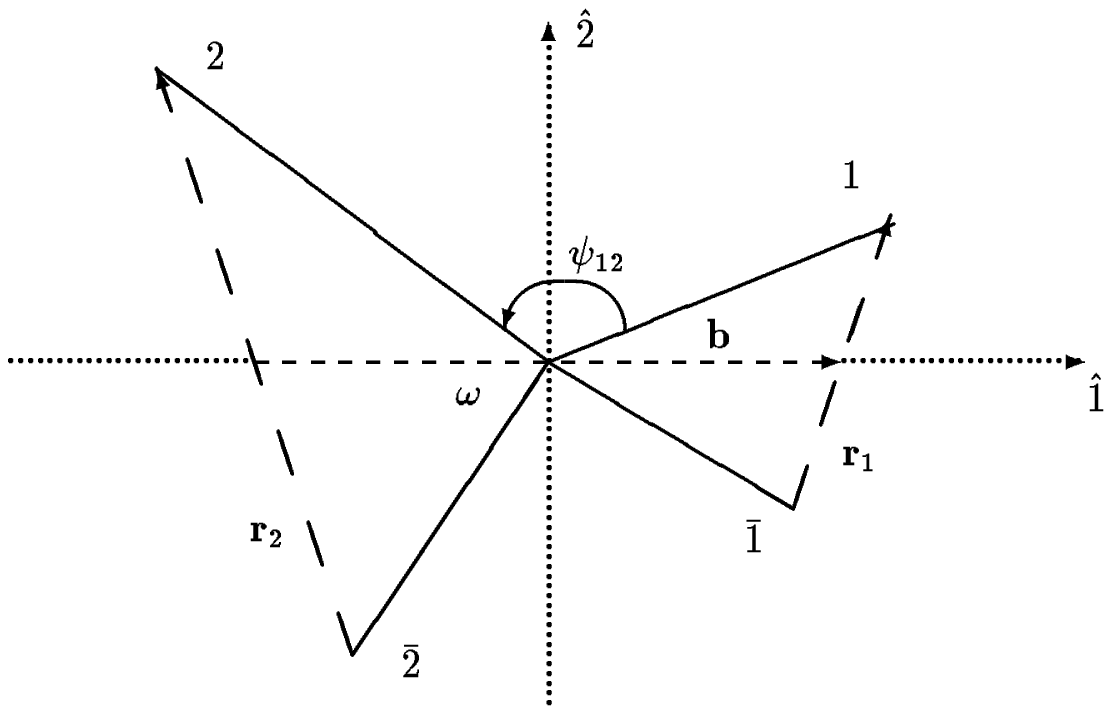}
$$
\caption{Space-time representation of the pyramids and their transverse 
projection. The sliding sides of the pyramids give  the domains $S_1$ and $S_2$ 
of the surface integrations. Note that the two Wilson loops are not parallel in 
the transverse plane.}
\label{surfaces}
\end{figure}

The various steps and approximations necessary to derive a tractable expression 
of $J$ have been developped in Ref.~\cite{dos94} to which we direct the interested 
reader. Here we only summarize these steps. First one transforms the line integrals 
appearing in the non abelian phases $W_{1}$ and $W_{2}$ into surface integrals. 
The manipulation of gauge invariant quantities leads to the introduction of a 
reference point $\omega$ in between the two surfaces. The surfaces to be considered 
are two pyramids, $S_1$ and $S_2$, with $\omega$ as apex and $\partial S_1$, 
$\partial S_2$ as respective basis. This is shown in Fig.~\ref{surfaces}. The surfaces 
are the world sheets of infinitely many gluons in the two-hadron state which 
interact via the correlator Eq.~(\ref{correlator}). In order to make the calculation 
practical, these interactions are truncated to fourth order in the field strengths. 
Because of the Gaussian process the terms factorize into products of integrals over 
the correlators linking the surfaces generated by $W_1$ and $W_2$. One gets
$$
J\approx{1\over 8N_C^2(N_C^2-1)}\left(
\int_{S_1}d\Sigma^{\mu\nu}(x)\int_{S_2}d\Sigma^{\rho\sigma}(y)
\langle g^2F^c_{\mu\nu}(x,\omega)F^c_{\rho\sigma}(y,\omega)\rangle_A\right)^2.
$$
Symbolically, this expression can be rearranged into a sum of four interaction-terms, 
$\chi$,
\begin{equation}\label{dipole-dipole}
J={1\over 8N_C^2(N_C^2-1)}\left({\langle g^2FF\rangle\over 12}\right)^2 \Big\{
\chi(q_1q_2)+\chi(\bar{q}_1\bar{q}_2)-\chi(q_1\bar{q}_2)-\chi(\bar{q}_1q_2)\Big\}^2,
\end{equation}
where each term has confining and non-confining parts from the basic gluon-gluon 
correlator in the vacuum,  
$\chi(qq)=\kappa\chi_c(qq)+(1-\kappa)\chi_{nc}(qq)$, which we now specify.

It is possible to choose functions $D$ and $D_1$ in a way that the natural fall ofF 
at large Euclidean distances does not impede a meaningful continuation to 
Minkowski space. A possible choice consistent with confinement and lattice 
computation~\cite{dig92} is
$$
D(z^2/a^2)={27\pi^4\over4}i\int {d^4k\over2\pi^4}{k^2\over(k^2-(3\pi/8)^2)^4}
e^{-ik\cdot z/a}.
$$
The function $D_1$ is a priori an independent one, however lattice simulation
indicates that the behavior of $D$ and $D_1$ is similar and we shall assume for 
simplicity that $D_1=D$.

\begin{figure}
$$
\epsfxsize=10cm\epsfbox[140 100 650 475]{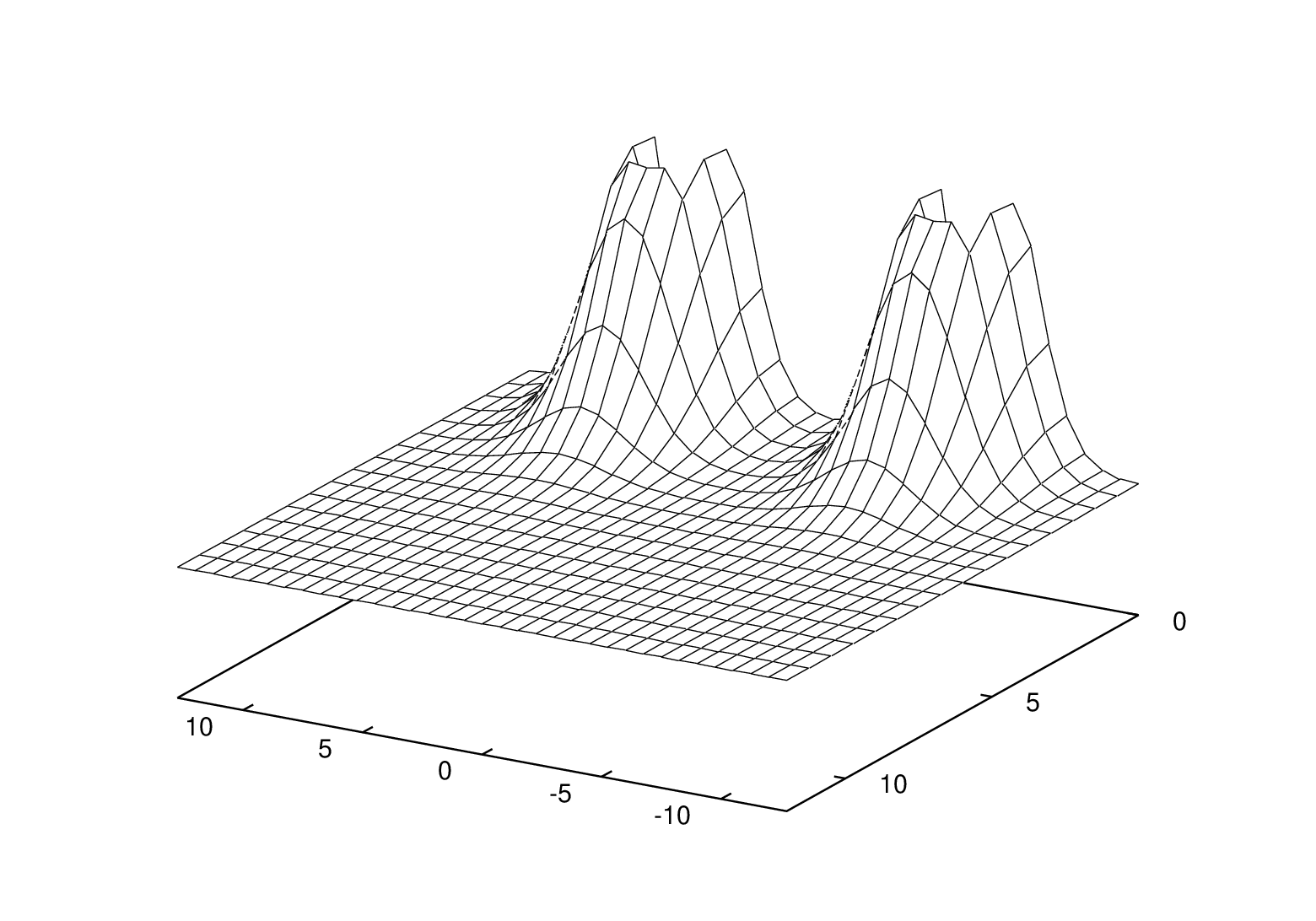}
\unitlength1cm
\begin{picture}(0,0)
\put(-1,.5){$y/a$}
\put(-9.5,.5){$x/a$}
\end{picture}
$$
$$
\epsfxsize=10cm\epsfbox[140 100 650 475]{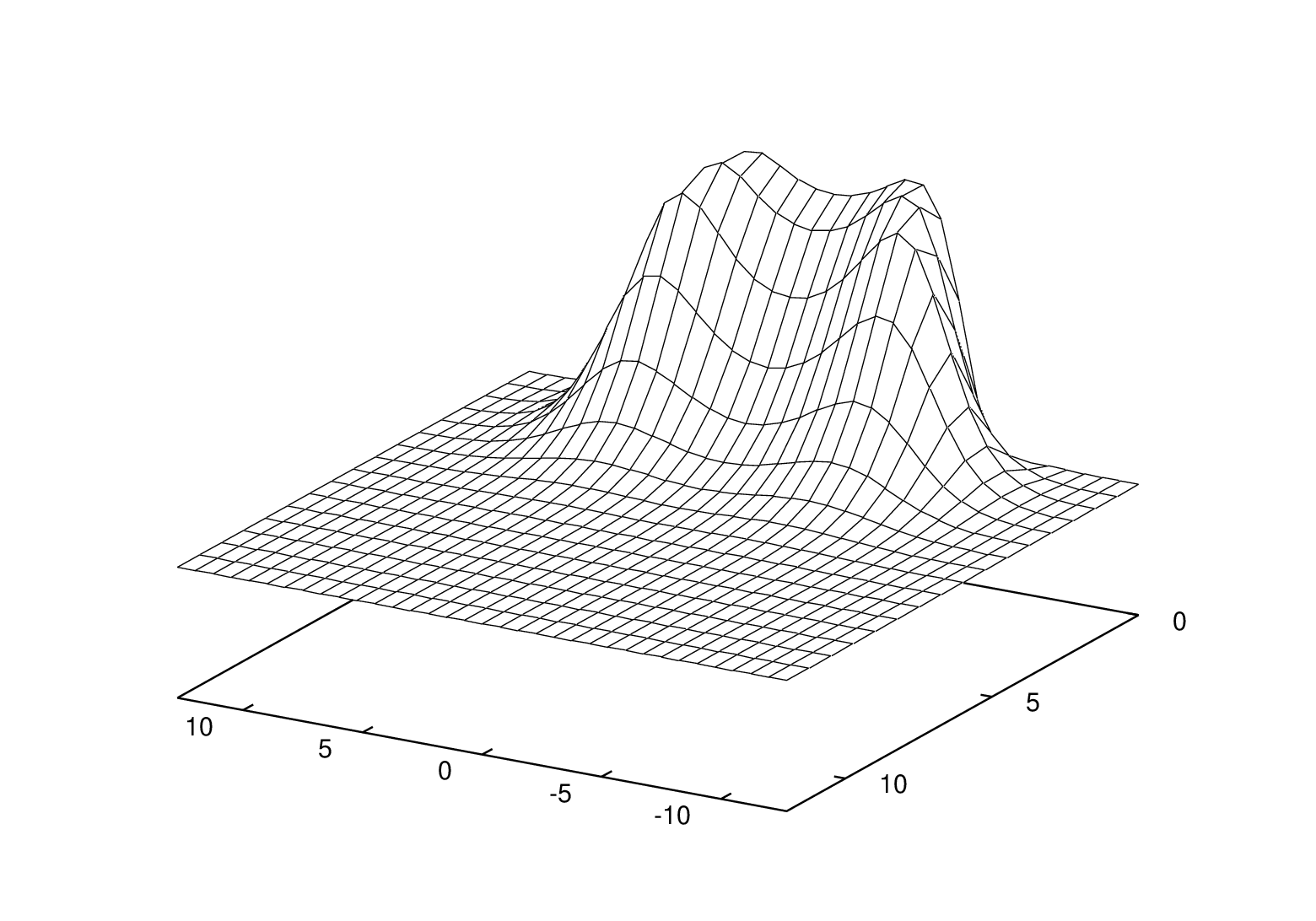}
\unitlength1cm
\begin{picture}(0,0)
\put(-1,.5){$y/a$}
\put(-9.5,.5){$x/a$}
\end{picture}
$$
\caption{(a) Color interaction amplitude Eq.~(\protect\ref{dipole-dipole}) for 
the non-confining case, $\kappa=0$, as a function of the impact position between 
the two dipoles. Dipole 1 has a transverse size $r_1=a$ and we sum over its 
orientation. Dipole 2 has a transverse size $r_2=12\,a$ and lies along the $x$-axis. 
(b) Color interaction amplitude for the confining case, $\kappa=1$.}
\label{color-interaction}
\end{figure}

For the non-confining part the two surface integrals can be performed and result 
in quark-quark interaction terms such as
$$
\chi_{nc}(q_1q_2)={8\over3}\left({|{\bf x}_1-{\bf x}_2|\over a}\right)^3\,
K_3\left({3\pi\over8}{|{\bf x}_1-{\bf x}_2|\over a}\right).
$$
Since nonperturbative gluon correlations are of size $a$, two color charges 
can only interact if their trajectories enter in a common domain of size $a$. 
A constituent quark picture arises where the elementary color charges are 
surrounded with gluon clouds. This can be seen in Fig.~\ref{color-interaction}(a) 
where the interaction amplitude $J$ for $\kappa=0$ between a dipole target 
of size $r_2=12\,a$ oriented along a given $x$-axis and a dipole probe of size 
$r_1=a$ is plotted as a function of the impact position. For simplicity we 
sum over the orientation of the probe. It turns out that with a physical 
correlation length around $0.3\,$fm a physical target has a size of 
$4\,a$--$5\,a$ so that constituents in the target are not as well separated 
as in Fig.~\ref{color-interaction}. 

For the confining part on the contrary the integrals have a  path dependence which 
is linked to the non-abelian nature of the confining term. Physically this means 
that the color dipoles connected by their strings interact as whole objects rather 
than as isolated endpoints. In space-time the integration over the surfaces 
is done with the reference point $\omega$ chosen in the most symmetrical way as 
shown in Fig.~\ref{surfaces}. The result depends only very weakly on the choice of the 
reference point and has the following form: 
\begin{eqnarray}
&&\chi_{c}(q_1q_2)={\pi\over2} \cos \psi_{12}\\
&&\left\{ {|{\bf x}_1-{\bf x}_{\omega}|\over a}\int_0^1d\alpha 
\left|{{\bf x}_1-{\bf x}_{\omega}-\alpha({\bf x}_2-{\bf x}_{\omega})\over a}\right|^2
K_2\left({3\pi\over8a}
|{\bf x}_1-{\bf x}_{\omega}-\alpha({\bf x}_2-{\bf x}_{\omega})|\right)
+(1\leftrightarrow2)\right\}.\nonumber
\end{eqnarray}
The angle $\psi_{12}$ denotes the angle between the vectors 
${\bf x}_1-{\bf x}_{\omega}$ and ${\bf x}_2-{\bf x}_{\omega}$. 
The amplitude in the string-string interaction picture, i.e. $\kappa=1$, is shown in 
Fig.~\ref{color-interaction}(b) with the same choice for the target and probe sizes 
as in the non-confining case. The interaction is non-zero whenever the probe is 
close to the line connecting the target quark and antiquark. 
We note that the string and constituent picture both differ from the optical 
droplet picture where the charge distribution form factor is responsible for 
the differential cross section. Via the wave functions of the valence quarks the 
geometrical sizes of the hadrons enter in the cross sections.

\subsection{From dipole-dipole to hadron-hadron cross section}\label{hadron-hadron}

A valence quark picture can be constructed from nonperturbative scattering 
amplitude of color dipoles with fixed lengths $r_1$ and $r_2$ by distributing 
the positions of the end-points of the strings according to a quantum mechanical 
wave function. Since for high-energy scattering the incoming particles propagate 
along the light cone, it is natural to choose light cone wave functions. The 
amplitude of the process can be written as~\cite{gun77}
\begin{equation}\label{amplitude1}
{\cal M}=2is\int d^2{\bf b} e^{-i{\bf \Delta}_T\cdot{\bf b}}
\int {dz_1d^2{\bf r}_1\over 4\pi}\psi^\dagger_V\psi_\gamma(z_1,{\bf r}_1)
\int {dz_2d^2{\bf r}_2\over 4\pi}|\psi_p(z_2,{\bf r}_2)|^2\,J(\{{\bf x}_i\}),
\end{equation}
where the index ``1'' refers to the photon or vector meson side whereas the 
index ``2'' is attached to the nucleon coordinates. (Conventions are fixed in 
Appendix~\ref{wf-computation}.) For simplicity, the nucleon is 
considered in a quark-diquark configuration. It has been shown in 
Ref.~\cite{dos94} that quark-diquark and three-quark pictures lead to 
similar predictions for diffractive scattering once the model parameters are 
adjusted to fit the proton-proton cross section. In the following, we also fix 
the parameters to fit proton-proton cross section and therefore we do not 
expect any significative dependence on the model of the proton. For 
$C=P=-1$ exchange, the quark-diquark is favoured since it suppresses the 
odderon contribution~\cite{rue96}. 

The $\psi$'s are the valence light cone wave functions of the corresponding hadrons. 
They are usually defined in momentum space where they describe the probability 
amplitudes to find in a hadron with  momentum $\{P^+,{\bf P}\}$\footnote{
Light cone coordinates are $P^{\pm}=(P^0\pm P^3)/\sqrt{2}$, 
${\bf P}={\bf P}_T=(P^1,P^2)$. A particle momentum is fully specified by the set 
$\{P^+,{\bf P}\}$ ($P^-=({\bf P}^2+M^2)/2P^+$ with $M$ the particle mass).}
and well defined angular momentum and flavor content  a quark and an antiquark
with momenta $\{zP^+,{\bf k}+z{\bf P}\}$ and $\{(1-z)P^+,-{\bf k}+(1-z){\bf P}\}$.
One crucial property of light cone wave functions is their dependence on $z$ and 
${\bf k}$ alone~\cite{bjo71,lep82}. This implies that upon Fourier transformation 
in the transverse plane, the relative coordinate of the $q \bar{q}$ pair, 
${\bf r}={\bf x}_q-{\bf x}_{\bar{q}}$, is easily separated from the position of the 
hadron ``center'', ${\bf X}=z{\bf x}_q+(1-z){\bf x}_{\bar{q}}$. In a similar way to 
non relativistic physics, the non trivial degree of freedom of the wave function 
can be isolated, the resulting transition matrix element is
$$
\langle z,{\bf x}_q,{\bf x}_{\bar{q}}|P^+,{\bf P}\rangle=\psi(z,{\bf r})e^{i{\bf P}\cdot{\bf X}}.
$$

In the amplitude given in Eq.~(\ref{amplitude1}) the impact parameter ${\bf b}$ 
denotes the transverse separation between hadron centers,
${\bf b}={\bf X}_1-{\bf X}_2$. The 
transverse positions of quarks are then given by
\begin{eqnarray*}
{\bf x}_1&=&{\bf x}_0+{\bf b}/2+(1-z_1){\bf r}_1,\\
{\bf x}_2&=&{\bf x}_0-{\bf b}/2+(1-z_2){\bf r}_2,\\
{\bf x}_0&=&({\bf X}_1+{\bf X}_2)/2,
\end{eqnarray*}
i.e. ${\bf x}_0$ is in the center of ${\bf X}_1$ and ${\bf X}_2$. Antiquark positions are 
related to quark positions as ${\bf x}_{\bar{\imath}}={\bf x}_i-{\bf r}_i$. We notice 
that the reference point $\omega$ chosen in Sec.~\ref{dipole-scattering} ``moves'' 
with respect to ${\bf x}_0$, 
${\bf x}_{\omega}={\bf x}_0-1/2[(z_1-1/2){\bf r}_1-(z_2-1/2){\bf r}_2]$. We have, 
however, checked that choosing $x_{0}$ rather than $\omega$ as the reference point 
for the computation of the loop-loop amplitude has negligible numerical effects.

In the present study, we are interested in electroproduction of different vector 
mesons under various kinematical conditions on a fixed proton target. It is therefore 
instructive to isolate the variable part of the amplitude associated with the photon and 
vector meson coordinates, $z_1,{\bf r}_1$, by integrating out the nucleon coordinates, 
$z_2,{\bf r}_2$. To this end we define
\begin{equation}\label{jproton}
J_p(z_1,{\bf r}_1,\Delta_T)=2\int_0^{+\infty} bdb\,2\pi J_0(\Delta_T b)
\int {dz_2d^2{\bf r}_2\over 4\pi}|\psi_p(z_2,{\bf r}_2)|^2 
J(b,z_1,{\bf r}_1,z_2,{\bf r}_2),
\end{equation}
so that the amplitude Eq.~(\ref{amplitude1}) is now written as 
\begin{equation}\label{amplitude2}
{\cal M}=is\int {dz_1d^2{\bf r}_1\over 4\pi}
\psi^\dagger_V\psi_\gamma(z_1,{\bf r}_1)\,J_p(z_1,{\bf r}_1,\Delta_T).
\end{equation}

The determination of parameters of the model of the stochastic vacuum can be 
made in different ways. We follow the method given in Ref.~\cite{dos94}. We use 
as input parameters the total proton-proton cross section at $\sqrt{s}=20\,$GeV, 
namely $\sigma_{pp}=35\,$mb and the slope $B_{pp}=11.5\,$GeV$^{-2}$ of the 
$p$-$p$ elastic cross section. From lattice simulations~\cite{dig92}, we take the 
mixing coefficient $\kappa=0.74$ and the curve relating gluon condensate and 
correlation length. The square of the proton quark-diquark wave function is 
taken in the simple form 
$$
|\psi(z,r)|^2=4\,\omega^2 \delta(z-1/2)e^{-\omega^2 r^2}.
$$ 
As ouput we obtain the gluon condensate $\langle g^2FF\rangle=2.49\,$GeV$^4$, 
the correlation length $a=0.346\,$fm and the proton transverse radius 
$R_{p}=1.51a=0.52\,$fm. The parameters are different from those of 
Ref.~\cite{dos94} where the influence of the non-confining term, $D_1$, in the 
correlator Eq.~(\ref{correlator}) was neglected, i.e. $D_1$ was set to 0. The 
proton size is smaller than the transverse rms charge radius, $R_T=0.68\,$fm. 
Sea-quark contributions may enter into the form factor increasing the charge 
radius compared to the radius of the valence quarks. The same sea quarks or 
extra color dipoles may also enter in the cross section at higher energies where 
it is increasing with energy. New lattice simulations\cite{dig96} try 
to isolate the perturbative contributions from the nonperturbative gluon 
fluctuations in a more precise way and may be incorporated  together with 
a better treatment of the perturbative two-gluon exchange.
\smallskip

\begin{figure}[ht] \leavevmode \centering
\fips{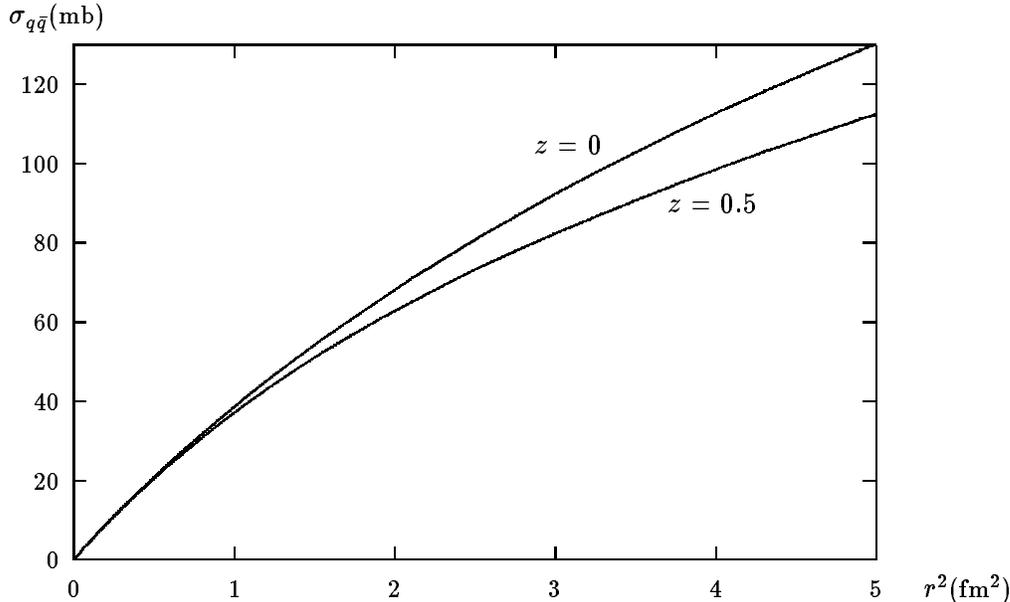}
\caption{Dipole-proton total cross section as a function of the dipole size for $z=0$ 
and $z=0.5$. The dependence in $z$ is rather marginal as it becomes noticeable  only 
for very large separation of the $q\bar{q}$ pair. The cross section behaves as 
$\sigma_{q\bar{q}}\propto r^n$ with $n=2$ for small extension and slowly decreasing 
at larger distances.}
\label{dipole-forward-amplitude}
\end{figure}

In Fig.~\ref{dipole-forward-amplitude}, we show the behavior of the function 
$$
J_p^{(0)}(z,r,\Delta_T=0)=\int_0^{2\pi}{d\theta\over2\pi}J_p(z,r,\theta,\Delta_T=0).
$$
It represents the total cross section of a $q\bar{q}$ dipole of fixed size $r$ 
averaged over its orientation. For varying dipole sizes the total cross section 
on the proton increases quadratically for small dipoles until a size of 
$r=1$--$2\,a$, then the increase continues but with a decreasing power. This 
feature is distinct to model of perturbative gluon exchange in Ref.~\cite{nik91} 
where this total cross section saturate at about twice the proton radius.

It is also instructive to consider the $\Delta_T$-fall off of the transition 
amplitude on the hadron radii and the correlation length $a$.
In Ref.~\cite{dos94}, the logarithmic slope $B$ of elastic cross sections has
been numerically parametrized as
$$
B\approx 1.56\,a^2+0.24\,(R_1^2+R_2^2).
$$
The large first term which is independent of the radii is specific for the model.

\section{Electroproduction of vector mesons}\label{electroproduction}

In this section, we describe the $q\bar{q}$ wave functions of the photon and 
vector meson which enter into  the expression of the production amplitude 
Eq.~(\ref{amplitude2}). Even at the qualitative level there are still large 
uncertainties about the correct dynamical description of hadrons. Indeed, one 
of the most interesting issues of current and future experiments is to shed light 
on the long distance properties of QCD. This evidently includes the unravelling 
of basic facts about hadronic wave functions. We shall show that the 
electroproduction of vector mesons is quite sensitive to their wave function.

For the time being, one has to make assumptions which influence the result as 
strongly as the dynamical features of the transition operator for diffraction 
given by the model of the stochastic vacuum. In the analysis of hadron-hadron 
scattering~\cite{dos94}, a simple transverse wave function for the $\pi$ meson
has been chosen 
\begin{equation}\label{gaussian}
\psi(r)=2\omega e^{-\omega^2 r^2/2}.
\end{equation}
Note that in this reference the dependence of the geometry of the loop-loop 
interaction on the respective light cone fractions of the quarks and antiquarks 
was not yet considered. Including this dependence and wave functions $\psi(z,r)$ 
with a reasonable behavior on the light cone momentum fraction $z$, we found 
similar results to those given in Ref.~\cite{dos94}. The transverse size of the 
studied hadrons still determines the size of hadron-hadron cross sections. We 
shall see in the following how this feature is modified when one is considering 
photon induced reaction.

\subsection{Photon wave function}

The $q\bar{q}$ wave function of the photon carries as labels the virtuality $Q^2$ 
and the polarization state $\lambda$ of the photon. It describes the probability 
amplitude to find a quark-antiquark pair inside the  photon with light cone 
fractions $(z,1-z)$ and  transverse separation ${\bf r}=(r\cos\theta,r\sin\theta)$. 
The $q\bar{q}$ state is in a configuration with  given flavor ($f,\bar{f}$) and 
helicities ($h,\bar{h}$). The color part of the wave function is treated separately 
and considered together with the Wilson loop in the way described in 
Sec.~\ref{dipole-scattering} and we are only left here with an overall factor 
$\sqrt{N_c}$. The photon couples to the electric charge of the quarks with
$e_f\delta_{f\!\bar{f}}$ where $e_f=2/3\,e$ or $-1/3\,e$ respectively. The 
helicity and spatial configuration part of the wave function looks different for 
various photon polarizations. It can be computed in light cone perturbation 
theory and  one has to lowest order (see the Appendix~\ref{wf-computation})
\begin{equation}\label{photon}
\psi_{\gamma(Q^2,\lambda)}=
\sqrt{N_c}\,e_f\delta_{f\!\bar{f}}\,\tilde{\psi}_{\gamma(Q^2,\lambda)},
\end{equation}
with
\begin{eqnarray*}
\tilde{\psi}_{\gamma(Q^2,0)}&=&-
\delta_{h,-\bar{h}}\,2z(1-z)Q {K_0(\varepsilon r)\over 2\pi},\\
\tilde{\psi}_{\gamma(Q^2,1)}&=&
\sqrt{2}\left(ie^{i\theta}\varepsilon\Big(z\delta_{h+}\delta_{\bar{h}-}
-(1-z)\delta_{h-}\delta_{\bar{h}+}\Big){K_1(\varepsilon r)\over 2\pi}
+m_{f}\delta_{h+}\delta_{\bar{h}+}{K_0(\varepsilon r)\over 2\pi}\right),\\
\tilde{\psi}_{\gamma(Q^2,-1)}&=&
\sqrt{2}\left(ie^{-i\theta}\varepsilon\Big((1-z)\delta_{h+}\delta_{\bar{h}-}
-z\delta_{h-}\delta_{\bar{h}+}\Big){K_1(\varepsilon r)\over 2\pi}
+m_{f}\delta_{h-}\delta_{\bar{h}-}{K_0(\varepsilon r)\over 2\pi}\right),
\end{eqnarray*}
where $\varepsilon=\sqrt{z(1-z)Q^2+m_{f}^2}$ and $m_f$ the current quark 
masses given in Table~\ref{meson-list} for the different flavors. $K_0$, $K_1$ 
are modified Bessel functions.

The longitudinal photon wave function is peaked around $z=1/2$, so that the 
longitudinal photon interacts like a small dipole, $r\sim 1/Q$, at large $Q^2$. 
On the contrary, the transverse photon is almost flat in $z$ so that, at large 
$Q^2$, it interacts partly like a small object for intermediate $z$ and partly 
like a large one for $z\sim m_f/Q$ when light quarks are involved. For heavy 
quarks, $c$, $b$, the inverse of the quark mass limits the photon extension. In 
electroproduction of vector mesons the effective dipole size is fixed by the 
overlap of wave functions of the photon with the vector meson. Due to the shape 
of the latter the small $z$ region is somewhat suppressed and the transverse 
region explored for $Q$ above $1$--$2\,$GeV is below or around $1\,$fm.
Lacking a better knowledge in the region of large transverse size, we use the 
above wave function. This may be tested by forthcoming experiments on the 
ratio of longitudinal to transverse cross sections. 

\subsection{Vector meson wave function}\label{hadronwf}

For the  hadron wave function, we use the same notation as described above for 
the photon. We have already taken care of color in the construction of the 
Wilson loops cf. Sec.~\ref{dipole-scattering}. For the flavor content we 
consider that the $\rho^0(770)$, $\omega(782)$, $\phi(1020)$ and $J/\psi$ 
mesons are respectively pure isospin 1, isospin 0, $s\bar{s}$ and $c\bar{c}$ 
vector mesons. This flavor part together with the configuration part
$\tilde{\psi}$ forms the wave function $\psi_V$  needed in Eq.~(\ref{amplitude2}).

Some information has been obtained on the spatial dependence of hadron wave 
functions. The first piece of information comes from long distance physics which 
various quark models describe successfully. These models tell us that a hadron at 
rest can be modelled with Gaussian wave functions as a sytem of  constituent 
quarks moving in a harmonic oscillator potential. To boost non  relativistic wave 
functions to a fast moving system is not a trivial step. Technically  the interplay 
between the transverse and longitudinal dynamics in light cone physics as well as 
the treatment of spin degrees of freedom remain to be understood. Within the 
model of the stochastic vacuum a nice quantitative description of hadron-hadron 
soft collisions has been obtained by disregarding spin and light cone fraction 
dependences. Therefore we assume that for soft collisions between large objects, a 
simple-minded description of hadrons  Eq.~(\ref{gaussian}) suffices. A smooth 
$z$-dependence on the light cone momentum fraction will not change this picture. 

The second piece of information comes from short distance physics and perturbative 
QCD supplemented by sum-rules, where some properties of the valence 
wave function at 0-transverse separation are known (in particular the end-point 
$z\to 0,1$ behavior can be analyzed). When wave functions at short distances are 
involved, e.g. in hard exclusive scatterings~\cite{lep80}, the value of the meson 
wave function at the origin determines the value of cross sections. The wave function 
at the origin  is related to the  leptonic decay width of the meson.

The above observations allow to make an ansatz for vector meson 
wave functions:
\begin{eqnarray}
\tilde{\psi}_{V(0)}&=&z(1-z){\delta_{h,-\bar{h}}\over\sqrt{2}}\,
{\sqrt{2}\pi f_V\over\sqrt{N_c}\hat{e}_V}f(z)e^{-{\omega^2 r^2\over2}},\nonumber\\
\label{vector-meson}
\tilde{\psi}_{V(1)}&=&\left\{{i\omega^2 re^{i\theta}\over M_V}
\Big(z\delta_{h+}\delta_{\bar{h}-}-(1-z)\delta_{h-}\delta_{\bar{h}+}\Big)
+{m_{f}\over M_V}\delta_{h+}\delta_{\bar{h}+}\right\}\,
{\sqrt{2}\pi f_V\over\sqrt{N_c}\hat{e}_V}f(z)e^{-{\omega^2 r^2\over2}},\\
\tilde{\psi}_{V(-1)}&=&\left\{{i\omega^2 re^{-i\theta}\over M_V}
\Big((1-z)\delta_{h+}\delta_{\bar{h}-}-z\delta_{h-}\delta_{\bar{h}+}\Big)
+{m_{f}\over M_V}\delta_{h-}\delta_{\bar{h}-}\right\}\,
{\sqrt{2}\pi f_V\over\sqrt{N_c}\hat{e}_V}f(z)e^{-{\omega^2 r^2\over2}}.\nonumber
\end{eqnarray}
This ansatz has the following properties. The main transverse  dependence 
$e^{-w^2r^2/2}$ and the function $f(z)$ are modeled in the way proposed by 
Wirbel, Stech and Bauer~\cite{wir85}
\begin{equation}\label{l-c-fraction-dep}
f(z)={\cal N}\sqrt{z(1-z))}e^{-M_V^2(z-1/2)^2/2\omega^2}.
\end{equation}
We use the same functional form for all vector mesons. The transverse size 
parameter $\omega$ is related to the vector meson radius, 
$\langle R_3^2\rangle^{1/2}=\langle x^2+y^2+z^2\rangle^{1/2}$, which for 
the Gaussian shape adopted is 
$\langle R_3^2\rangle^{1/2}=\smash{\sqrt{3/2}}/2\omega$. This quantity is 
presumably not very different from the electromagnetic radius which 
unfortunatly is unknown for vector mesons. The way out in the quark model is 
to fix $\omega$ and ${\cal N}$ by the normalization and the $e^+ e^-$ decay 
width (see Appendix~\ref{wf-parameters}). We draw attention to the fact that 
applying this procedure to the above parametrization of the wave function 
leads to different sets of parameters, $\{\omega,{\cal N}\}$, for longitudinal 
and transversal mesons. We find radii in this way which are reasonable in the 
whole family of vector mesons. This is a welcome property because, as we shall 
see in Sec.~\ref{numerics}, the vector meson transverse size is one of the 
important ingredients determining the cross sections in the intermediate 
$Q^2$-range. The above form is written to have explicitly the correct value 
of the wave function at the origin, i.e. the $q\bar{q}$-state fulfills the equation 
$$
\langle0|J^{\mu}(0)|V(q,\lambda)\rangle=ef_V M_V\varepsilon^{\mu}(q,\lambda),
$$
with $f_V$ the meson decay constant. In Eq.(\ref{vector-meson}), $\hat{e}_V$ 
is the mean quark charge in the meson state in units of the proton charge 
(see Table~\ref{meson-list} in Appendix~\ref{wf-parameters}). The helicity 
dependence of the wave functions is modeled after the perturbative 
$\gamma\to q\bar{q}$ transition.

\subsection{Cross sections}

Let us collect the formula necessary to compute cross sections. 
It is convenient to expand the quantity $J_p$ in Eq.(\ref{jproton}) 
in terms of $L_z$-eigenfunctions 
$$
J_p(z_1,r_1,\theta_1,\Delta)=\sum_m e^{im\theta_1} J^{(m)}_p(z_1,r_1,\Delta),
$$
where thanks to the periodicity of $J_p$ only even $m$ are present. It follows 
that in the reaction studied, only transverse to transverse and longitudinal to 
longitudinal transitions are expected. From the explicit form of the wave 
function, Eqs.~(\ref{photon}) and~(\ref{vector-meson}), there is a possibility of 
helicity change by two units in the process. We have, however, observed that 
the corresponding contribution to the cross section is smaller than 2\% in the 
whole $Q^2$-range. We disregard this contribution.   

In the following, we distinguish transverse and longitudinal cross sections which, 
in the present conventions, are 
\begin{eqnarray}
{d\sigma_L\over dt}&=&{1\over16\pi}\left|\int{dz_1r_1dr_1\over 2}
\psi^\dagger_{V(0)}\psi_{\gamma(0)}(z_1,r_1)\,J^{(0)}_p(z_1,r_1,\Delta)\right|^2
\nonumber\\
\label{diff-cross-sec}
{d\sigma_T\over dt}&=&{1\over16\pi}\left|\int {dz_1r_1dr_1\over 2}
\psi^\dagger_{V(1)}\psi_{\gamma(1)}(z_1,r_1)\,J^{(0)}_p(z_1,r_1,\Delta)\right|^2,
\end{eqnarray}
where the average over proton helicities is understood. The combination 
$\psi^\dagger_{V(\lambda)}\psi_{\gamma(\lambda)}(z_1,r_1)$ are computed 
by multiplying Eq.~(\ref{photon}) by Eq.~(\ref{vector-meson}) supplemented by 
the flavor part described in Sec.~\ref{hadronwf} giving
\begin{eqnarray}
\psi^\dagger_{V(0)}\psi_{\gamma(0)}&=&-ef_Vz_1(1-z_1)f(z_1)e^{-\omega^2r_1^2/2}\, 
2z_1(1-z_1)QK_0(\varepsilon r_1)\nonumber\\
\label{photon-meson}
\psi^\dagger_{V(1)}\psi_{\gamma(1)}&=&ef_Vf(z_1)e^{-\omega^2r_1^2/2}\left\{
{\omega^2\varepsilon r_1\over M_V}\Big[z_1^2+(1-z_1)^2\Big]K_1(\varepsilon r_1)
+{m_{f}^2\over M_V}K_0(\varepsilon r_1)\right\}
\end{eqnarray}

Experimentally, the differential cross sections Eq.~(\ref{diff-cross-sec}) are difficult 
to measure. On the one hand, the separation of transverse and longitudinal cross 
section is not easily done, on the other hand accurate data exist only for $t$-integrated 
cross section. Nevertheless, some results have been obtained for 
$$
{d\sigma_{\rm exp}\over dt}=\epsilon{d\sigma_L\over dt}+{d\sigma_T\over dt},
$$
as a function of $\Delta_T^2$. $\epsilon$ is the rate of longitudinally polarized 
photons which depends on the lepton scattering angle, $\theta_e$, and the photon 
energy, $\nu$. In the proton rest frame
$$
\epsilon=\left[1+2\left(1+\nu^2/Q^2\right)\tan^2(\theta_e/2)\right]^{-1}.
$$
We shall also compare our theoretical results with the integrated cross section 
$\sigma_{\rm exp}=\epsilon\sigma_L+\sigma_T$, for various $Q^2$. By 
analyzing the vector meson decay, it is possible to check the validity of 
$s$-channel helicity conservation or assuming helicity conservation to deduce 
$R=\sigma_L/\sigma_T$.

\section{Numerical investigation}\label{numerics}

\subsection{General results}\label{general-results}

Before entering in a detailed analysis, let us discuss qualitatively the 
$Q^2$-dependence of differential cross section expected from 
Eq.~(\ref{diff-cross-sec}). The quantity $J^{(0)}_p(z_1,r_1,t)$ is power behaved 
in $r_1$ and decreases exponentially with $|t|$. It depends only weakly on 
$z_1$. At fixed $t$, $J^{(0)}_p\propto r_1^{\alpha}$ with $\alpha=2$ at 
small $r_1$ and $\alpha$ slowly decreasing for $r_1\gtrsim a$ (see 
Fig.~\ref{dipole-forward-amplitude}). Given this behavior, we also need an 
estimate of the effective size of the photon-vector meson overlap after 
integration over $z_1$ in Eq.~(\ref{photon-meson}). One can anticipate two 
extreme regimes. At small $Q^2$, the effective size is driven by the meson wave 
function so that $d\sigma_L\propto Q^2$ and $d\sigma_T$ is constant.
At large $Q^2$, the effective size is given by the photon wave function alone,
which leads to $d\sigma_L\propto Q^{-6}$ and $d\sigma_T\propto Q^{-8}$.

\begin{figure}
$$
\fips{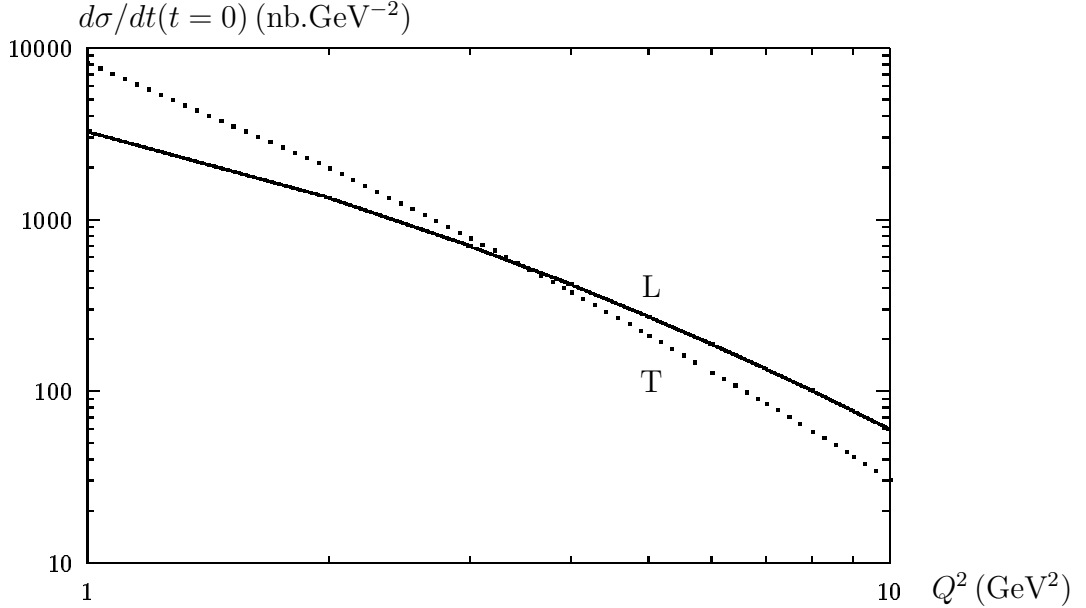}
\setlength{\unitlength}{0.240900pt}
\begin{picture}(0,0)
\put(150,23){\makebox(0,0){$Q^2\,$(GeV$^2$)}}
\put(-1300,920){\makebox(0,0)[l]{$d\sigma/dt(t=0)\,$(nb.GeV$^{-2}$)}}
\put(-400,500){\makebox(0,0){L}}
\put(-400,350){\makebox(0,0){T}}
\end{picture}
$$
\caption{$d\sigma/dt(t=0)$ for $\rho$-production as a function of $Q^2$ for 
the longitudinal (full) and transverse (dots) cross section. The effective power of 
the fall off with $Q$ is increased by 2 units in the range 1--10$\,$GeV$^2$ but 
the asymptotic behavior is only reached in the 10--100$\,$GeV$^2$}\label{dsigma-dt}
\end{figure}

These asymptotic behaviors of cross section are rather model-independent. Most 
of the current experimental results, however, are below the asymptotic large 
$Q^2$-region. The theoretical behavior of cross sections in our calculation shows 
a much more specific dependence on $Q^2$ in this region. 
In Fig.~\ref{dsigma-dt} we show the logarithmic $\rho$-production cross section 
$\log d\sigma/dt(t=0)$ as a function of $Q^2$. The effective power $n$ of the 
fall off $d\sigma/dt(t=0)\propto Q^{-n}$ is about 2.5 and 4 at 
$Q^2=1\,$GeV$^2$ for the longitudinal and transversal cross section respectively 
and this power increases to about 4.5 and 6 respectively at $Q^2=10\,$GeV$^2$. 
Thus the asymptotic regime starts only above that value.

As we have emphasized the result depends on the wave function chosen to represent 
the vector meson state. Without changing the  parametrization, this can be seen 
by varying the size parameter $\omega$ of the vector meson wave function.
With the value of the wave function at the origin kept fixed we observe that 
decreasing $\omega$ by 5\% increases  the cross section at $Q^2=1\,$GeV$^2$
by about 20\%. This modification becomes less than 3\% at $Q^2=10\,$GeV$^2$ 
because the vector meson size is less important for a small $q$-$\bar q$
state in the photon. One can also think of changing the light cone distribution 
of the quark and antiquark in the meson wave function. An indication 
of the resulting modification is given by changing the factor 
$\sqrt{z(1-z)}\to [z(1-z)]^{3/2}$ in the parametrization of $f(z)$ in 
Eq.~(\ref{l-c-fraction-dep}). This leads to about 30\% decrease of the cross 
section in the whole $Q^2$ range examined.

\begin{figure}
$$
\fips{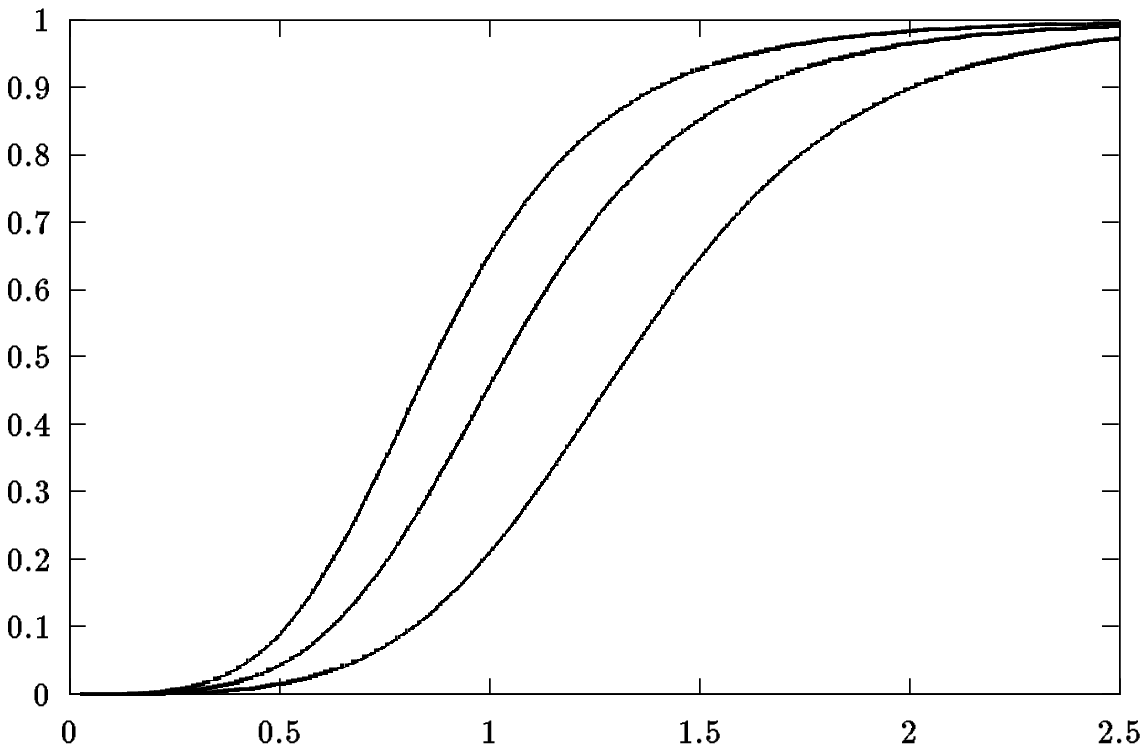}
\setlength{\unitlength}{0.240900pt}
\begin{picture}(0,0)
\put(150,23){\makebox(0,0){$r_1^{\rm cut}\,$(fm)}}
\put(-1300,920){\makebox(0,0)[l]{
$\left.\frac{d\sigma_T}{dt}\right|_{t=0}(r_1^{\rm cut})
/\left.\frac{d\sigma_T}{dt}\right|_{t=0}(r_1^{\rm cut}=\infty)$}}
\put(-910,450){\makebox(0,0){10}}
\put(-720,450){\makebox(0,0){4}}
\put(-580,450){\makebox(0,0){1}}
\end{picture}
$$
\vskip 5mm
$$
\fips{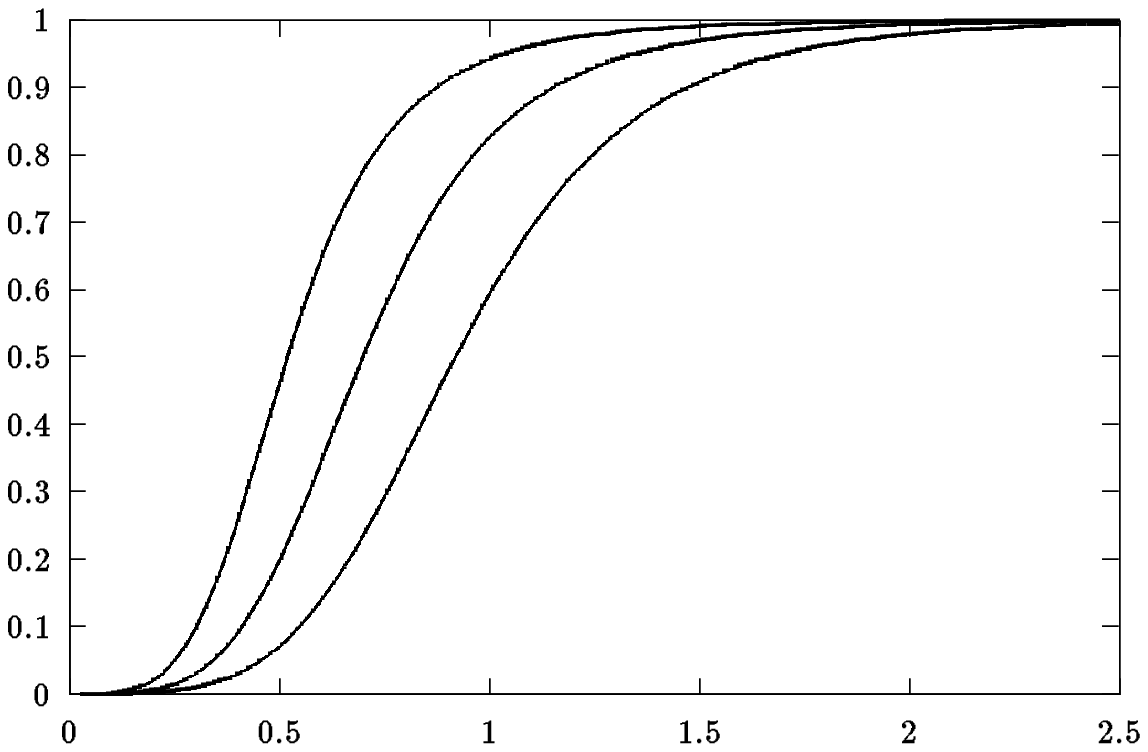}
\setlength{\unitlength}{0.240900pt}
\begin{picture}(0,0)
\put(150,23){\makebox(0,0){$r_1^{\rm cut}\,$(fm)}}
\put(-1300,920){\makebox(0,0)[l]{
$\left.\frac{d\sigma_L}{dt}\right|_{t=0}(r_1^{\rm cut})
/\left.\frac{d\sigma_L}{dt}\right|_{t=0}(r_1^{\rm cut}=\infty)$}}
\put(-1070,450){\makebox(0,0){10}}
\put(-880,450){\makebox(0,0){4}}
\put(-770,450){\makebox(0,0){1}}
\end{picture}
$$
\caption{(a) The contribution to the transverse differential cross section of 
$\rho$-production with respect to the radial parameter of the photon and 
vector meson side. The integration over the transverse distance $r_1$ is 
taken up to a cut-off $r_1^{\rm cut}$ for several values of the photon 
virtuality, $Q^2=1,\,4$ and $10\,\rm GeV^2$. The transverse differential 
cross section at $t=0$ is plotted as a function of $r_1^{\rm cut}$. It is 
normalized to its value with the cut-off removed. (b) Same as (a) for longitudinal 
polarization. With increasing $Q^2$ the cross section is dominated by short 
transverse distances in both cases but the saturation occurs earlier for the 
longitudinal cross section.}
\label{dsdt-rcut}
\end{figure}

In Fig.~\ref{dsdt-rcut}, we show the importance of the transversal extension of 
the virtual photon. We introduce a transversal cut-off in the cross section, i.e. 
we cut the $r_1$-integration in Eq.~(\ref{amplitude2}) at a fixed value 
$r_1\leq r_1^{\rm cut}$. As expected the $q\bar{q}$-wavefunction in 
a transversal photon extends to much larger values of $r_1$, e.g. at 
$Q^2=4\,$GeV$^2$ more than 50\% of the cross section comes from transverse 
separations bigger than 1$\,$fm. In the longitudinal case, the contribution of the
region with transverse separations $r_1>1\,$fm drops down to 15\% at 
$Q^2=4\,$GeV$^2$.

\subsection{$\rho$, $\omega$, $\phi$-production}

\begin{figure}
$$
\fips{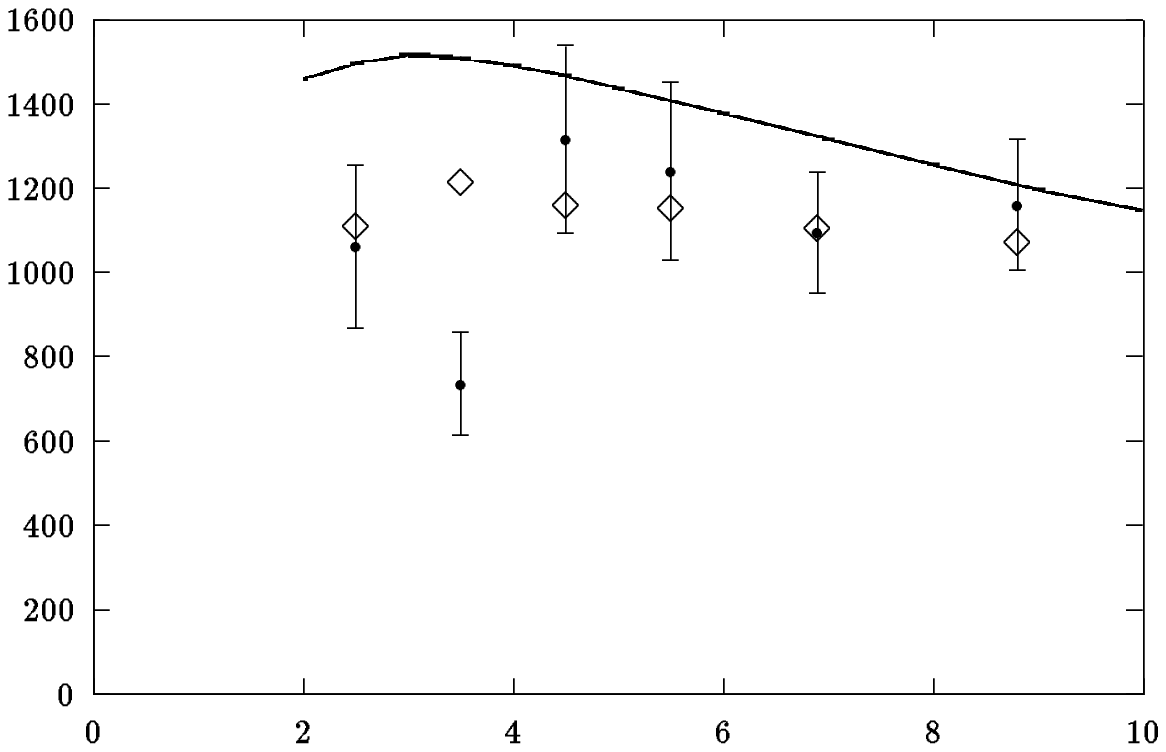}
\setlength{\unitlength}{0.240900pt}
\begin{picture}(0,0)
\put(150,23){\makebox(0,0){$Q^2\,$(GeV$^2$)}}
\put(-1300,920){\makebox(0,0)[l]{$Q^4\,\sigma\,$(nb.GeV$^{4}$)}}
\put(-150,310){\makebox(0,0)[r]{$Q^4(\sigma_L+\sigma_T)$}}
\put(-122.0,310.0){\rule[-0.200pt]{15.899pt}{0.800pt}}
\put(-150,220){\makebox(0,0)[r]{$Q^4(\epsilon\sigma_L+\sigma_T)|_{\rm model}$}}
\put(-100,220){\raisebox{-.8pt}{\makebox(0,0){$\Diamond$}}}
\put(-150,130){\makebox(0,0)[r]{$Q^4(\epsilon\sigma_L+\sigma_T)|_{\rm NMC}$}}
\put(-100,130){\circle*{12}}
\end{picture}
$$
\caption{The scaled cross section $Q^4\,\sigma(Q^2)$ for $\rho$-production 
in nb.GeV$^4$. The circles are the NMC-results\protect\cite{nmcrho} and the 
diamonds represent our prediction for the quantity 
$Q^4\,(\epsilon\sigma_L+\sigma_T)$ with the experimental polarization 
rate of NMC.}\label{integrated-cross-section}
\end{figure}

In the range from 1 to 10$\,$GeV$^2$, the $1/Q^4$ behaviour of the production
cross section observed by EMC and NMC is very well reproduced by our calculations. 
Besides the $Q^2$-dependence, also the absolute values of the cross section 
are reproduced. It should be noted that within our model we have introduced no 
new parameters and the parameters underlying the interaction on the quark-gluon 
level are determined by soft high-energy proton-(anti)proton scattering. We 
show in Fig.~\ref{integrated-cross-section} NMC deuteron data~\cite{nmcrho} 
together with our prediction for $Q^4(\sigma_L+\sigma_T)$. We notice that 
this theoretical quantity is not the actually measured cross section due to the 
polarization rate of the photon, $\epsilon\ne 1$. Therefore we also show with 
diamonds the quantity $Q^4(\epsilon\sigma_L+\sigma_T)$ at the $Q^2$-points 
of NMC using their value of $\epsilon(Q^2)$.

In our approach, the approximate $1/Q^4$ behavior is due to a combination 
of different fall offs  in $\sigma_L$ and $\sigma_T$ which themselves come from 
the interplay between the size dependence of the dipole-proton cross section and 
the effective size of the photon-meson overlap. This is very different from the 
dynamics which occur in quark-quark scattering which leads however to a 
similar $Q^2$-dependence. As explained above the asymptotic behaviour 
for large $Q^2$ is just governed by the dipole size of the virtual photon and thus 
model independent. 

\begin{figure}
$$
\fips{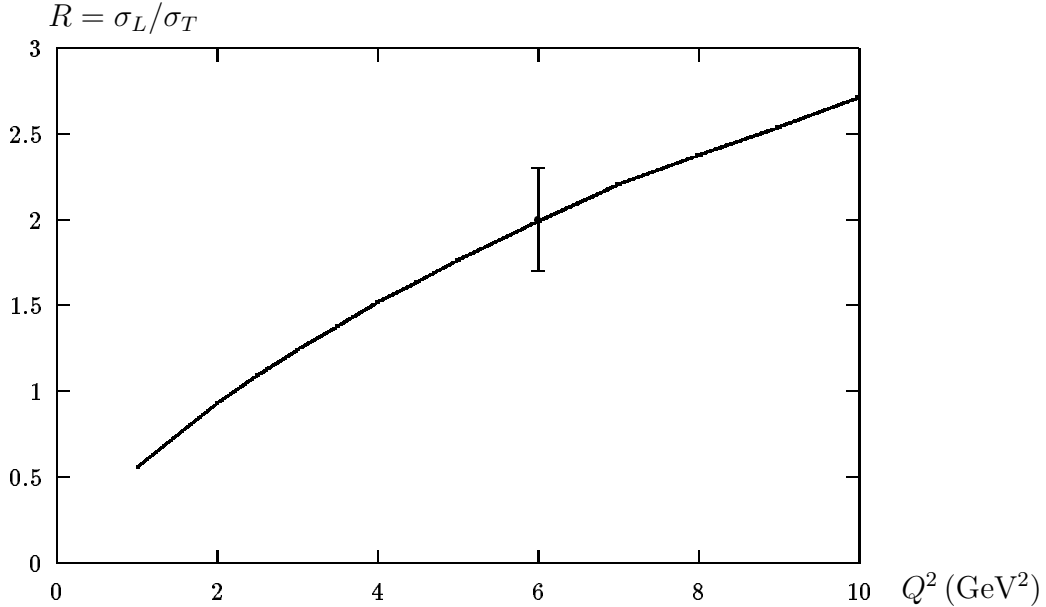}
\setlength{\unitlength}{0.240900pt}
\begin{picture}(0,0)
\put(150,23){\makebox(0,0){$Q^2\,$(GeV$^2$)}}
\put(-1300,920){\makebox(0,0)[l]{$R=\sigma_L/\sigma_T$}}
\end{picture}
$$
\caption{The ratio for longitudinal to transverse cross section for 
$\rho$-production. The data point is from Ref.~\protect\cite{nmcrho}. 
Other data compare well within errors but they either are far outside
the 10--$20\,$GeV range or have large errorbars.}
\label{ratio}
\end{figure}

A possible way to distinguish both approaches would be a precise measurement 
of the $Q^2$-behavior of the ratio of longitudinal to transverse cross sections 
$R(Q^2)=\sigma_L/\sigma_T$. We plot this ratio in Fig.~\ref{ratio}. For large 
$Q^2$, $R\propto Q^2$ but this behavior is not yet reached in the intermediate 
range where $R$ grows slower than $Q^2$. Here we expect a rather different 
behavior in our model compared to other models. 

\begin{figure}
$$
\fips{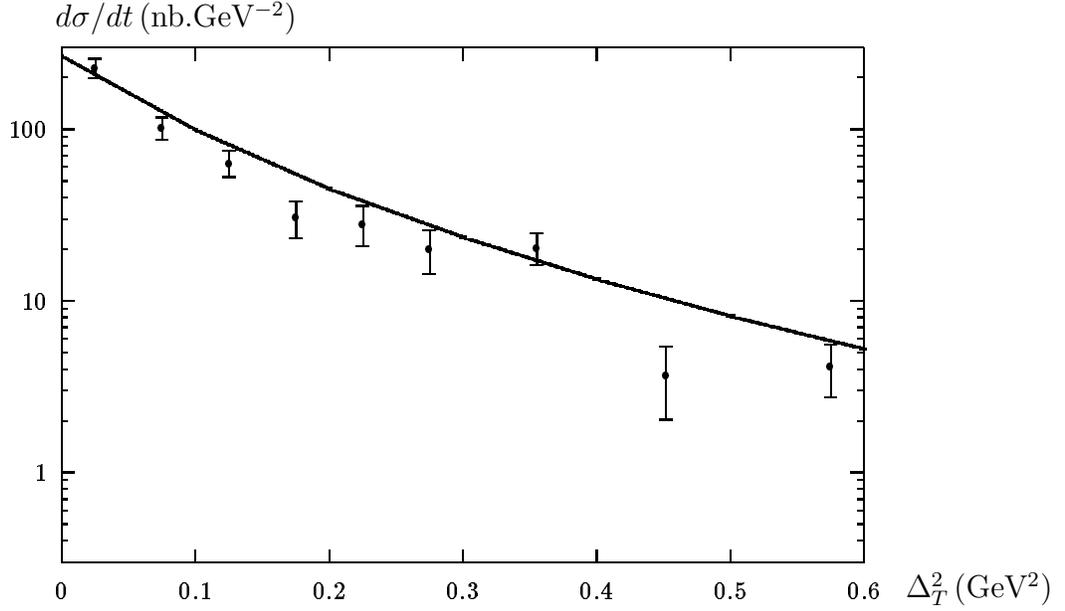}
\setlength{\unitlength}{0.240900pt}
\begin{picture}(0,0)
\put(150,23){\makebox(0,0){$\Delta_T^2\,$(GeV$^2$)}}
\put(-1300,920){\makebox(0,0)[l]{$d\sigma/dt\,$(nb.GeV$^{-2}$)}}
\end{picture}
$$
\caption{
The differential cross section, $d\sigma/dt(\Delta_T^2)$, for $\gamma^*+p\to\rho+p$ 
at $Q^2=6\,$GeV$^2$. Data from Ref.~\protect\cite{nmcrho}.}
\label{pt2-dependence}
\end{figure}

Another important check is provided by looking at the $p_T$-dependence of 
the differential cross section. We show our result for 
$\epsilon d\sigma_L/dt+d\sigma_T/dt$ versus $\Delta_T^2$ at 
6$\,$GeV$^2$ and compare in Fig.~\ref{pt2-dependence} to the NMC points 
for the deuteron outside of the coherent production region~\cite{nmcrho}. 
Notice that the $t$-dependence of our computation is not fully exponential.

\begin{figure}
$$
\fips{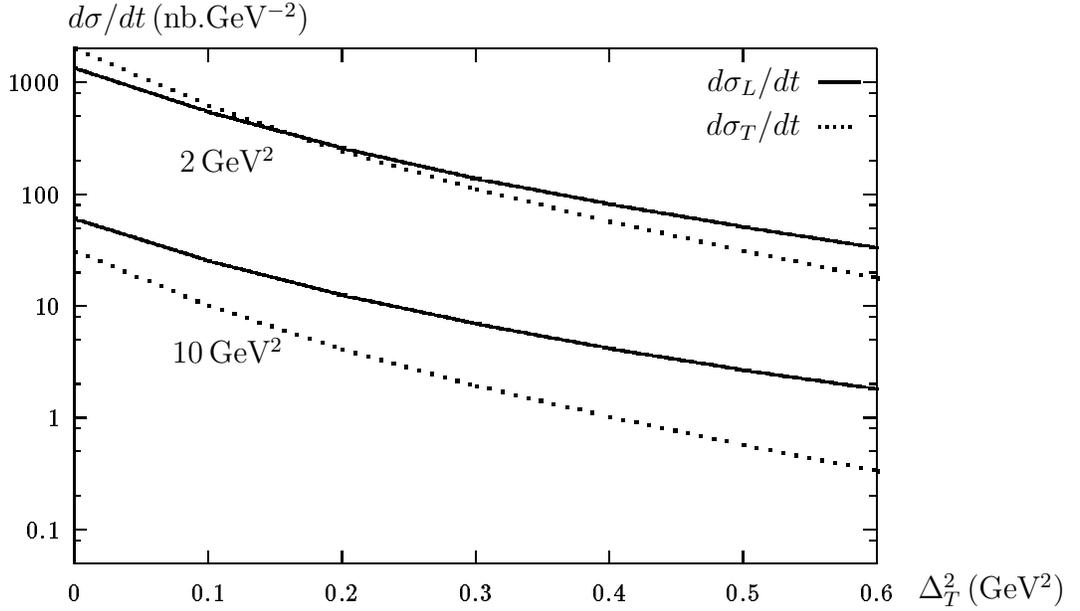}
\setlength{\unitlength}{0.240900pt}
\begin{picture}(0,0)
\put(150,23){\makebox(0,0){$\Delta_T^2\,$(GeV$^2$)}}
\put(-1300,920){\makebox(0,0)[l]{$d\sigma/dt\,$(nb.GeV$^{-2}$)}}
\put(-150,820){\makebox(0,0)[r]{$d\sigma_L/dt$}}
\put(-122.0,820.0){\rule[-0.200pt]{15.899pt}{0.800pt}}
\put(-150,750){\makebox(0,0)[r]{$d\sigma_T/dt$}}
\multiput(-122.0,750.0)(12,0){5}{\rule[-0.200pt]{0.8pt}{0.8pt}}
\put(-1050,700){\makebox(0,0){$2\,$GeV$^2$}}
\put(-1050,400){\makebox(0,0){$10\,$GeV$^2$}}
\end{picture}
$$
\caption{
The differential cross section for $\rho$-production, $d\sigma/dt(\Delta_T^2)$, 
for a longitudinal photon (full lines) and a transverse one (dotted lines) at $Q^2=2$ 
and $10\,$GeV$^2$.}\label{pt2-dep-Q2}
\end{figure}

There is a non trivial $Q^2$-dependence of the slope $B$ in our model which is 
due to a decreasing transverse region probed by the slowly shrinking size of the 
photon as $Q^2$ grows. We show this in Fig.~\ref{pt2-dep-Q2}.
\smallskip

For the production of the $\omega$-meson, we expect a wave function very similar 
to the one of the $\rho$ and correspondingly the ratio should be determined 
through the flavour factor
$$
f_{\omega}^2/f_{\rho}^2\approx 9\%.
$$
Indeed, this is observed.

The situation is more complex for the production of heavier vector mesons 
where the different quark content has several consequences. The direct 
appearance of a mass term in the meson wave function gives an additionnal 
component in the overlap Eq.~(\ref{photon-meson}) and also modifies 
the photon extension parameter $\varepsilon^2=z(1-z)Q^2+m_f^2$. The 
quark content also influences both the transversal size of the vector meson 
and its momentum fraction distribution $f(z)$. Whereas the first effects are 
quite easily controlable, nothing precise is known about the quantitative 
effect of a heavier mass on the distribution $f(z)$. Qualitatively, one expects 
that the distribution becomes more peaked at $z=1/2$ as the mass of the 
constituents increases. 

\begin{figure}
$$
\fips{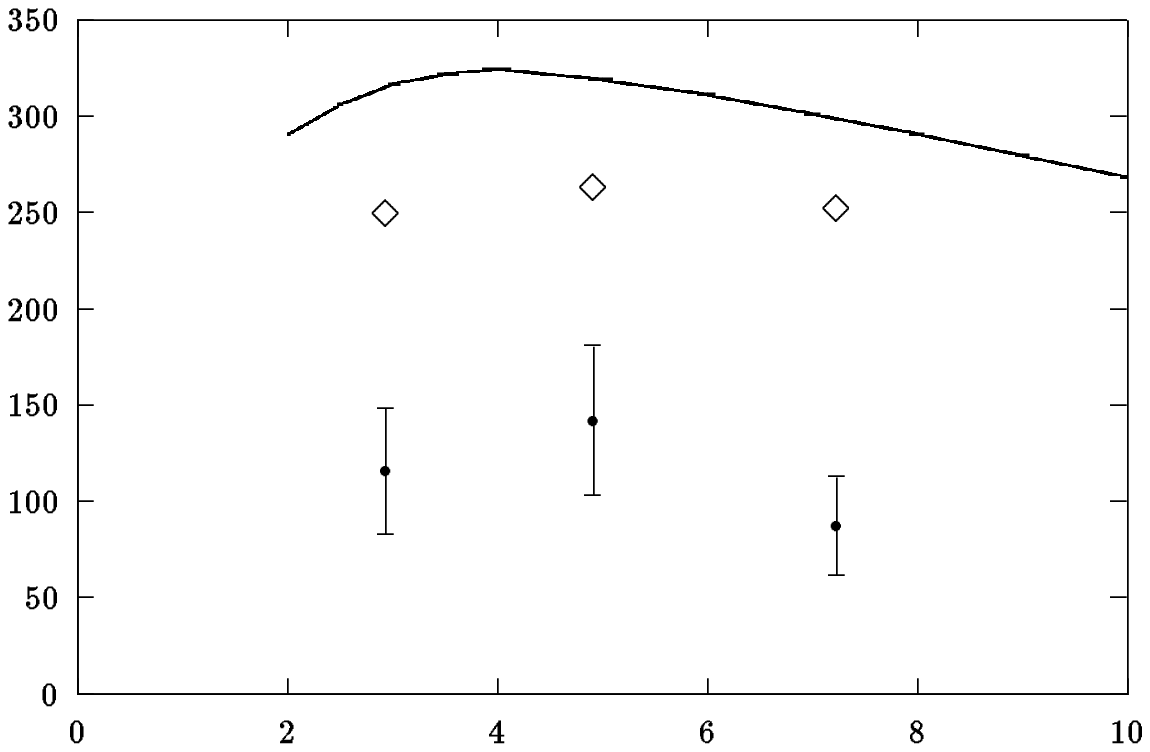}
\setlength{\unitlength}{0.240900pt}
\begin{picture}(0,0)
\put(150,23){\makebox(0,0){$Q^2\,$(GeV$^2$)}}
\put(-1300,920){\makebox(0,0)[l]{$Q^4\,\sigma\,$(nb.GeV$^{4}$)}}
\put(-150,550){\makebox(0,0)[r]{$Q^4(\sigma_L+\sigma_T)$}}
\put(-122.0,550.0){\rule[-0.200pt]{15.899pt}{0.800pt}}
\put(-150,480){\makebox(0,0)[r]{$Q^4(\epsilon\sigma_L+\sigma_T)|_{\rm model}$}}
\put(-100,480){\raisebox{-.8pt}{\makebox(0,0){$\Diamond$}}}
\put(-150,130){\makebox(0,0)[r]{$Q^4(\epsilon\sigma_L+\sigma_T)|_{\rm NMC}$}}
\put(-100,130){\circle*{12}}
\end{picture}
$$
\caption{$\phi$-production cross section compared to NMC 
data~\protect\cite{nmcrho}. In our model the difference between theoretical and 
experimental result is attributed to wave function effect.}\label{phi-cross-section}
\end{figure}

For $\phi$-meson, a smaller transversal extension than in the $\rho$-meson 
is expected due to the heavier $s$-quark mass. The smaller size of the 
$\phi$-meson reduces the $\phi$-production cross section at low 
$Q^2$ values beyond the flavour factor
$$
f_{\phi}^2/f_{\rho}^2\approx 27\%.
$$
Such an effect is also observed in the difference between pion-nucleon and 
kaon-nucleon scattering in the model~\cite{dos94}. It should diminish with 
increasing $Q^2$ in electroproduction since then the amplitude is less
and less determined by the extension of the produced meson, but rather
by the virtual photon. This effect may have been observed in the ZEUS 
data~\cite{zeuphi}. Also the change of the longitudinal distribution in $z$
between $\phi$ and $\rho$-meson may influence the cross section 
independently of $Q^2$. This difference is absent with our choice of 
distribution for the $\phi$-meson which is numerically the same as for the 
$\rho$-meson. Our resulting theoretical cross section for the 
$\phi$-production reproduce the $Q^2$ dependence of the NMC data, but 
its absolute value is practically a factor 2 too large (see 
Fig.~\ref{phi-cross-section}). 

The increase of the longitudinal to transverse ratio, $R(Q^2)$, looks the 
same as for the $\rho$-meson, its overall magnitude being just reduced by 
about 20\%. The $p_T$-dependence also exhibits a similar pattern with a small 
broadening of the diffraction peak as $Q^2$ is increased.

\subsection{$J/\psi$-production}

For heavier quark pairs, the large quark mass leads to more dramatic 
modifications. Let us first notice that in our model we assume that the quarks 
move on lightlike trajectories. This can only be the case at energies far above 
$2\,m_f$. Therefore a center-of-mass energy bigger than 10$\,$GeV is necessary in 
the $J/\psi$-case. At these energies a moderate energy dependence is observed 
which is not contained in our model. In return, the large quark mass provides a hard 
scale so that also photoproduction data are accessible within our perturbative 
treatment of the photon. Strictly speaking, the difference 
$t_0=-M_p^2(Q^2+M_V^2)^2/s^2$, between $t$ and $-\Delta_T^2$, leads to 
a phase space threshold, $e^{Bt_0}$, for the $J/\psi$-production in the energy 
range we are considering. At $\sqrt{s}=15\,$GeV, it is easy to see that this 
threshold effect is only sizeable for large $Q^2$, e.g. for $Q^2=10\,$GeV$^2$ 
and $B=5-10\,$GeV$^{-2}$ one gets $e^{Bt_0}=0.97-0.94$. We shall disregard 
this factor in the following. 

The discussion of the asymptotic regime given in Sec.~\ref{general-results} can 
be refined in the presence of a large quark mass. 
Let us reexamine the short distance regime in the presence of the mass-terms 
in Eq.~(\ref{photon-meson}). To simplify further, we temporarily assume a 
simple non-relativistic form for the distribution, i.e. $f(z)\propto\delta(z-1/2)$ 
and for consistency take $m_c=M_J/2$. We consider the domain of large enough 
$\varepsilon^2=m_c^2+Q^2/4=(M_J^2+Q^2)/4$, where we can approximate 
$\exp(-\omega^2r^2/2)\approx1$ and 
$J_p^{(0)}(z_1=1/2,r_1,\Delta=0)\approx Cr_1^2$. One gets
\begin{eqnarray*}
{d\sigma_L\over dt}(t=0)&=&\alpha_{\rm em}(8f_JC)^2{Q^2\over(M_J^2+Q^2)^4}
\\
{d\sigma_T\over dt}(t=0)&=&\alpha_{\rm em}(8f_JC)^2
{M_J^2\over(M_J^2+Q^2)^4}
\left({M_J^2+8\omega^2\over M_J^2+2\omega^2}\right)^2.
\end{eqnarray*}
From these expressions one sees that the relevant scale for $J/\psi$-production 
is $M_J^2+Q^2$ rather than $Q^2$. As in the light quark case, the longitudinal 
cross section is expected to dominate at large $Q^2$, namely the quantity 
$(M_J^2+8\omega^2)^2/(M_J^2+2\omega^2)^2\approx1.4$ leads to a ratio 
$R\approx 0.7\,Q^2/M_J^2$. The differential cross section is expected to fall off 
as $d\sigma/dt\propto(1+\epsilon R)(M_J^2+Q^2)^{-4}$. Experimentally 
a $Q^2$-dependence such as $(M_J^2+Q^2)^{-n}$ is observed with $n$ 
around 2 but the accuracy of the data is not sufficient to exclude a complete 
short distance fall off. Let us stress that an accurate test of this power-law 
is a necessary prerequisite to understand the physics at work.

\begin{figure}
$$
\fips{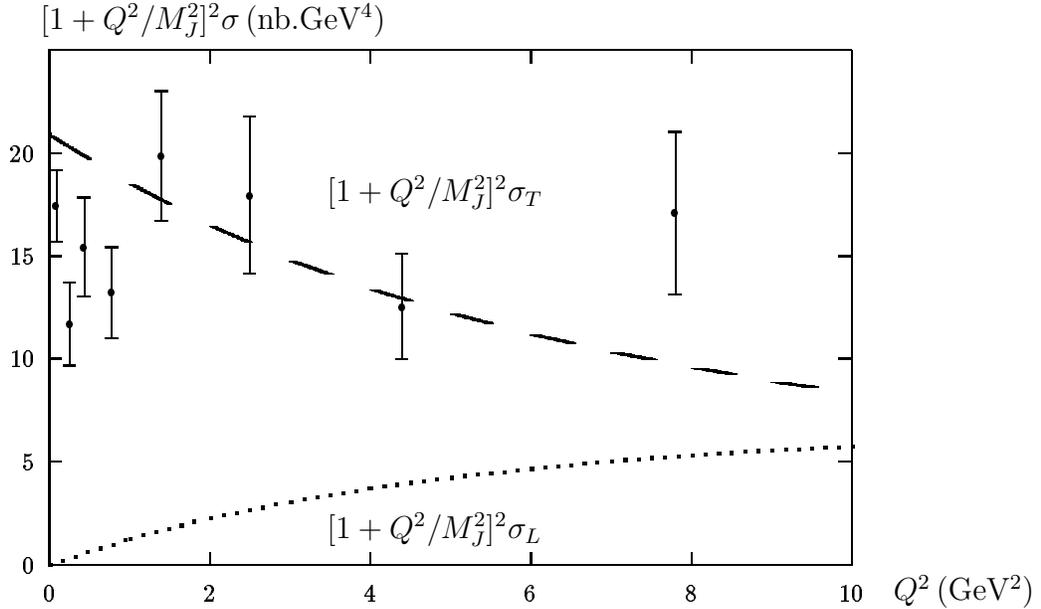}
\setlength{\unitlength}{0.240900pt}
\begin{picture}(0,0)
\put(150,23){\makebox(0,0){$Q^2\,$(GeV$^2$)}}
\put(-1300,920){\makebox(0,0)[l]{$[1+Q^2/M_J^2]^2\sigma\,$(nb.GeV$^4$)}}
\put(-850,120){\makebox(0,0)[l]{$[1+Q^2/M_J^2]^2\sigma_L$}}
\put(-850,650){\makebox(0,0)[l]{$[1+Q^2/M_J^2]^2\sigma_T$}}
\end{picture}
$$
\caption{$J/\psi$-production cross section for longitudinal (dots) and 
transverse (dashes) polarizations. To compare with EMC-data~\protect\cite{emcpsi}, 
one has to combine these two cross sections into $\sigma=\epsilon\sigma_L+\sigma_T$ 
with the polarization rate measured by EMC, $\epsilon\approx 0.7$.}
\label{psi-cross-section}
\end{figure}

In Fig.~\ref{psi-cross-section}, we plot $(1+Q^2/M_J^2)^2\,\sigma$, i.e. the cross 
section rescaled by the observed data fall off, for the transversal and longitudinal 
cross section separately together with the data recorded by EMC in the energy 
range $\sqrt{s}=$10--20$\,$GeV\cite{emcpsi}. The photoproduction comes out 
fairly in the energy range $\sqrt{s}=$10--20$\,$GeV, where several 
measurements have been performed leading to a production cross section 
between 10 and 20$\,$nb. As could be guessed from the study of the asymptotic 
behavior above, the shape of the $z$-distribution and the value of the charm 
quark mass determine the size of the cross section. Changing the quark mass by 
5\% would lead to a 20\% change in the cross section at $Q^2=0$ and to a 10\% 
change at $Q^2=10\,$GeV$^2$ respectively. As $Q^2$ increases, our expectation 
follows qualitatively the pattern depicted for the short distance regime although 
quantitatively the intermediate transverse distance somewhat contributes to give 
a fall off flatter than the short distance one.

\begin{figure}
$$
\fips{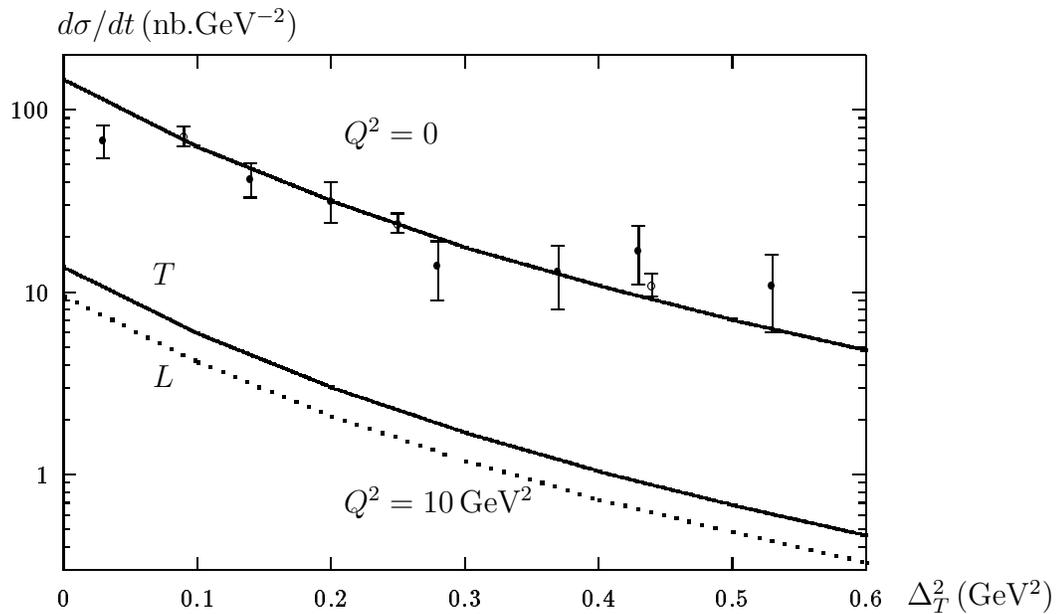}
\setlength{\unitlength}{0.240900pt}
\begin{picture}(0,0)
\put(150,23){\makebox(0,0){$\Delta_T^2\,$(GeV$^2$)}}
\put(-1300,920){\makebox(0,0)[l]{$d\sigma/dt\,$(nb.GeV$^{-2}$)}}
\put(-850,170){\makebox(0,0)[l]{$Q^2=10\,$GeV$^2$}}
\put(-850,750){\makebox(0,0)[l]{$Q^2=0$}}
\put(-1150,530){\makebox(0,0)[l]{$T$}}
\put(-1150,370){\makebox(0,0)[l]{$L$}}
\end{picture}
$$
\caption{
$d\sigma/dt(\Delta_T^2)$ for $J/\psi$-production at $Q^2=0$ and 
$10\,$GeV$^2$. The upper curve is our prediction for the photoproduction 
differential cross section. It can be compared to the measurement of 
Ref.~\protect\cite{e401} (filled circles) and to the extrapolation to $Q^2=0$ 
of the EMC data~\protect\cite{emcpsi}. Similar data have been measured by 
NMC~\protect\cite{nmcpsi} at $Q^2=1.5\,$GeV$^2$. Also shown are the 
differential cross sections at $Q^2=10\,$GeV$^2$ for longitudinal (dotted line) 
and transverse (lower full line) photons. The dependence of the slope on $Q^2$ 
and polarization is marginal in the $J/\psi$-case.}
\label{psi-pt2-dep}
\end{figure}

We next turn to the study of the $p_T$-dependence shown in Fig.~\ref{psi-pt2-dep}.
We find a good agreement with the photoproduction measurement~\cite{e401} and 
with the extrapolation of EMC (open circles). We also note that, contrarily to the 
``large'' hadron case, there is practically no $Q^2$-dependence of the $p_T$-fall 
off in the $J/\psi$-case.

\section{Conclusion}

We have calculated the longitudinal and transversal differential cross
sections for diffractive production of  $\rho$, $\omega$, $\phi$ and 
$J/\psi$-mesons in the range of $2\,$GeV$^2 \leq  Q^2 \leq 10\,$GeV$^2$. 
The hadronic part of our calculation is based on a model for nonperturbative 
QCD, the model of the stochastic vacuum \cite{dos87}. The parameters of 
the model which gives a unified description of low energy and soft high-energy 
scattering phenomena can be obtained from a variety of sources: hadron 
spectroscopy, QCD sum rules, high-energy proton-proton scattering and 
lattice calculations of the fundamental gluon field correlator. 
We have used a consistent parameter set very similar to the one used for
hadron-hadron scattering in Ref.~\cite{dos94}. The virtual photon and hadron 
wave functions are light cone wave functions motivated by perturbation 
theory for the photon and relativistic quark models for the hadrons, we thus 
do not have adjustable parameters. A specific feature of electroproduction is 
the dependence of the cross sections on the photon virtuality $Q^2$ which is 
reproduced by the model almost perfectly. Even at $Q^2=10\,$GeV$^2$ the 
cross sections have not yet reached their asymptotic $1/Q^6$ behaviour. Our 
calculation is consistent with the observed ratio of longitudinal to transverse 
cross sections. Here precise data at different values of $Q^2$ could 
discriminate between different models. In our model, we also can calculate 
the dependence on the tranverse momentum transfer and it nicely reproduces 
the available data.

Depending on photon polarization, the $Q^2$ range 2--$10\,$GeV$^2$ corresponds 
to an effective transverse $q\bar{q}$ size lying between $0.5\,$fm and $1.2\,$fm. 
This region just interpolates between the short distance domain and normal hadron 
diameters, thus allowing a natural extension of the phenomenology of 
hadron-hadron scattering where the model has been applied originally. 
Model dependent features of the light cone wave functions enter into 
the magnitude of the production cross section. A good experimental separation 
of $\sigma_L$ and $\sigma_T$ can help to obtain a real breakthrough in our 
understanding of diffractive electroproduction, since the physics of the $q\bar{q}$ 
pair state is so much different for both photon polarizations. The extension to 
photoproduction and low $Q^2$ electroproduction necessitates a modification 
of the simple perturbative $q\bar{q}$ wavefunction in the photon.

The model of the stochastic vacuum cannot predict the behaviour of the 
cross sections as a function of the cm energy $\sqrt{s}$. If all parameters are 
fixed, it yields constant cross sections. We therefore have confined ourselves 
in this paper to the energy range $10\,$GeV$\leq\sqrt{s} \leq 20\,$GeV. 
The absolute values for the cross sections are correctly reproduced for 
$\rho$, $\omega$ and $J/\psi$-production. Our results for $\phi$-production 
are about a factor 2 larger than the NMC data, we can, however, explain 
the $Q^2$-dependence of the ratio of $\phi$ to $\rho$ production as observed 
by ZEUS~\cite{zeuphi}. 

In the present model the energy dependence of the total hadronic cross section 
and the slope of the elastic cross section can be obtained consistently by 
increasing the hadron radii (slightly) with energy. Due to the expected 
difference of the energy dependence of the perturbative and nonperturbative 
contributions (hard and soft pomeron) one has also to take into account 
perturbative contributions if one wants to obtain a realistic model of the 
energy dependence of diffractive electroproduction. This will be done in a 
forthcoming work.

\acknowledgments

We are indebted to Andrzej Sandacz for providing us the NMC data and 
making clear related issues. We have benefited from discussions with Michael 
Rueter. 

\appendix

\section{Photon wave function computation}\label{wf-computation}

The photon wave function in the framework of light cone perturbation theory is 
discussed in~\cite{bjo71}. We compute it using the rules and conventions given 
in~\cite{lep80}. For a photon with momentum 
$q=[q^+,q^-=-Q^2/2q^+,{\bf q}={\bf 0}]$, one multiplies
\begin{itemize}
\item a color factor $\sqrt{N_c}$,
\item a flavor part $e_f\delta_{f\bar{f}}$,
\item a spinor term $\bar{u}(zq^+,{\bf k},h)\,\varepsilon_{\mu}(q,\lambda)
\gamma^{\mu} v((1-z)q^+,-{\bf k},\bar{h})$,
\item a factor $(\sqrt{2}zq^+)^{-1/2}(\sqrt{2}(1-z)q^+)^{-1/2}$ for the quark and 
antiquark lines,
\item a light cone energy denominator 
$-\sqrt{2}q^+\left[Q^2+{{\bf k}^2+m^2\over z(1-z)}\right]^{-1}$.
\end{itemize}
The polarization vectors of the photon are $\varepsilon(q,0)=[q^+/Q,Q/2q^+,{\bf 0}]$ 
and $\varepsilon(q,\pm 1)=[0,0,-1/\sqrt{2},\mp i/\sqrt{2}]$. The spinor matrix element 
between infinite-momentum-frame helicity eigenspinors~\cite{lep80} are 
\begin{eqnarray*}
\bar{u}\gamma^+ v&=&2\sqrt{z(1-z)}q^+\,\delta_{h,-\bar{h}},\\
\bar{u}\gamma^- v&=&-{{\bf k}^2+m^2\over\sqrt{z(1-z)}q^+}\,\delta_{h,-\bar{h}},\\
\bar{u}\gamma^i v&=&
{(1-2z)k^i\mp i\epsilon^{ij3}k^j\over\sqrt{z(1-z)}}\,\delta_{h,-\bar{h}}
\mp m{\delta^{i1}\mp i\delta^{i2}\over\sqrt{z(1-z)}}\,\delta_{h,\bar{h}}.
\end{eqnarray*}
In the last line $i=1,2$ and $\mp$ stands for a minus sign if $h=+1/2$ and for a plus 
sign when $h=-1/2$.

By taking the Fourier transform 
$$
\psi(z,{\bf r})=\int {d^2{\bf k}\over (2\pi)^2} e^{i{\bf k}\cdot{\bf r}}\psi(z,{\bf k}),
$$
one gets the expressions given in Eq.~(\ref{photon}). In the longitudinal case there is 
an additional $\delta^{(2)}({\bf r})$ which one can drop because the color interaction 
vanishes at 0 transverse distance.
\smallskip

It is of course possible to obtain a wave function description in covariant approach 
and we want here to give the steps necessary to get the photon wave function. First, 
it is important to notice that a photon-quark-antiquark coupling in a Feynman 
graph can be interpreted in term of a photon wave function in light cone perturbation 
theory if the $x^+$-ordering is $\gamma\to q\bar{q}$. This is the case in the formal 
limit $q^+\to+\infty$ where this ordering survives. 

The first step is the evaluation of the $k^-$ integral which leads for asymptotic 
$q^+$ to
$$
{dk^+d^2{\bf k}\over (2\pi)^4}\int_{-\infty}^{+\infty}dk^-{f(k^+,k^-,{\bf k})\over
(k^2-m^2+i\varepsilon)((k-q)^2-m^2+i\varepsilon)}\sim
{dzd^2{\bf k}\over16\pi^3}{if(k^+,0^-,{\bf k})\over {\bf k}^2+m^2-z(1-z)q^2},
$$
provided $f(k^-=0)$ is finite and non-zero. The numerator, 
$N=i(k\slash+m)(-ie\varepsilon\slash_\lambda) i(k\slash-q\slash+m)$, is
\begin{eqnarray*}
N&=&ie\left\{ k\cdot \varepsilon\,(2k\slash-q\slash)-k\cdot (k-q)\,\varepsilon\slash+
i\epsilon_{\alpha\mu\nu\rho}\gamma_5\gamma^{\alpha}k^{\mu}\varepsilon^{\nu}
q^{\rho}+m(2k\cdot\varepsilon-\varepsilon\slash q\slash+m\varepsilon\slash)\right\}\\
&\approx&ieq^+\gamma^-\left\{{\delta_{\lambda 0}\over Q}[z(1-z)q^2+{\bf k}^2+m^2]
+(1-2z){\bf k}\cdot{\bf \varepsilon}_T+i\gamma_5\epsilon^{ij3} k^i\varepsilon^j_T
+m\varepsilon\slash_T\right\}.
\end{eqnarray*}
The wave function is obtained by taking the helicity matrix element 
$\bar{w}(h)N w(-\bar{h})/\sqrt{2}q^+$ with $w(1/2)=(1/\sqrt{2},0,-1/\sqrt{2},0)$ 
and $w(-1/2)=(0,1/\sqrt{2},0,1/\sqrt{2})$~\cite{lep80}.

\section{Hadron wave function parameters}\label{wf-parameters}

The value of the wave function at the origin is related to the meson leptonic decay 
constant
\begin{equation}
\langle0|J^{\mu}(0)|V(q,\lambda)\rangle=ef_V M_V\varepsilon^{\mu}(q,\lambda),
\end{equation}
which appears in the expression of the vector meson $e^+e^-$ width
$$
\Gamma(V\to e^+e^-)={4\pi\alpha^2\over 3M_V}f_V^2.
$$
In the parametrization Eq.~(\ref{vector-meson}), this constraint leads to the 
determination of the parameter $\cal N$. The fixing condition, which depends on 
the meson helicity, $\lambda$, is
\begin{equation}\label{origin}
1=\int_0^1 dz\,z(1-z)f_L(z)=\int_0^1 dz\,
{2[z^2+(1-z)^2]\omega_T^2+m^2\over 2M_V^2z(1-z)}f_T(z),
\end{equation}
with $\hat{e}_V$ the effective quark charge in the meson $V$ expressed in units 
of the proton charge (see Table~\ref{meson-list}). 

The normalization condition is 
\begin{equation}
\langle V(q',\lambda')|V(q,\lambda)\rangle=(2\pi)^3 2q^+\delta(q^+-q'^+)
\delta^{(2)}({\bf q}-{\bf q'})\delta_{\lambda\lambda'},
\end{equation}
which leads to the relation
\begin{equation}\label{normalization}
\omega_{\lambda}={\pi f_V\over\sqrt{2N_c}\hat{e}_V}\sqrt{I_{\lambda}},
\end{equation}
where
\begin{eqnarray*}
I_L&=&\int_0^1 dz\,z^2(1-z)^2f_L^2(z),\\
I_T&=&\int_0^1 dz\,{[z^2+(1-z)^2]\omega_T^2+m^2\over M_V^2}f_T^2(z).
\end{eqnarray*}
$\omega_{\lambda}$ and ${\cal N}_{\lambda}$ are therefore defined by a system 
of implicite equations, Eq.~(\ref{origin}) and Eq.~(\ref{normalization}). Solutions 
are listed in Table~\ref{meson-list} together with the corresponding root mean 
square radius $R=\langle R_3^2\rangle^{1/2}=\smash{\sqrt{3/2}}/2\omega$.

\begin{table}
\begin{tabular}{lcdddcccc}
$V(M_V)$&$\hat{e}_V$&$\Gamma$&$f_V$&$\omega_L$&$R_L$&$\omega_T$&$R_T$\\
\phantom{$V$}(GeV)&&keV&MeV&GeV&fm&GeV&fm\\
\hline
$\rho\,(770)$&$1/\sqrt{2}$&6.7(7)&153.&.33&.37&.22&.55\\
$\omega\,(782)$&$1/3\sqrt{2}$&0.60&45.8&.30&.40&.21&.58\\
$\phi\,(1019)$&1/3&1.37&79.1&.37&.33&.26&.46\\
$J/\psi\,(3097)$&2/3&5.2(6)&270.&.68&.18&.57&.21\\
\end{tabular}
\caption{\label{meson-list}
{\small Vector meson characteristics. The quark masses considered are $m_u=m_d=0$,
$m_s=0.15\,$GeV, $m_c=1.3\,$GeV.}}
\end{table}

\end{document}